\documentclass[a4paper,11pt]{article}
\pdfoutput=1 

\usepackage{jheppub} 
\usepackage{adjustbox}
\usepackage{subfig}
\usepackage{floatrow}
\usepackage{array}
\usepackage{multirow}
\usepackage{arydshln}
\newcolumntype{C}[1]{>{\centering\arraybackslash}p{#1}}
\newfloatcommand{capbtabbox}{table}[][\FBwidth]

\newcommand{\call}{\mbox{${\cal L}$}}
\newcommand\pt{P_{\scriptscriptstyle T}}
\newcommand\kwl{\kappa^{\scriptscriptstyle{(W)}}_{\scriptscriptstyle{L}}}
\newcommand\kwr{\kappa^{\scriptscriptstyle{(W)}}_{\scriptscriptstyle{R}}}
\newcommand\kzl{\kappa^{\scriptscriptstyle{(Z)}}_{\scriptscriptstyle{L}}}
\newcommand\kzr{\kappa^{\scriptscriptstyle{(Z)}}_{\scriptscriptstyle{R}}}
\newcommand\khl{\kappa^{\scriptscriptstyle{(H)}}_{\scriptscriptstyle{L}}}
\newcommand\khr{\kappa^{\scriptscriptstyle{(H)}}_{\scriptscriptstyle{R}}}
\newcommand\kwlr{\kappa^{\scriptscriptstyle{(W)}}_{\scriptscriptstyle{L/R}}}
\newcommand\kzlr{\kappa^{\scriptscriptstyle{(Z)}}_{\scriptscriptstyle{L/R}}}
\newcommand\khlr{\kappa^{\scriptscriptstyle{(H)}}_{\scriptscriptstyle{L/R}}}

\title{NLO QCD corrections to single vector-like top partner production in association with top quark at the LHC}

 \author{Mohamed Sadek Zidi}
 \affiliation{LPTh, Department of Physics, University of Jijel,\\B. P. 98 Ouled Aissa, 18000 Jijel, Algeria}

\emailAdd{mohamed.sadek.zidi@gmail.com}

\abstract{The single vector-like top partner hadroproduction in association with top quark in the five and the six-flavour schemes at the LHC is investigated. 
The inclusive cross section and the differential distributions are evaluated for many benchmark scenarios at leading and next-to-leading orders, both for fixed-order and fixed-order matched to parton shower. The associated Higgs production with a top quark pair, where the final state arises from the decay of the vector-like top partner into a Higgs boson and a top quark, is examined.
}

\keywords{Vector-like quarks, NLO QCD corrections, parton shower, 6F-scheme}

\begin{document} 
\maketitle
\flushbottom
\section{Introduction}
\label{sec1}
The Standard Model (SM) is a successful theory of particle physics. All the particles predicted by this theory were observed at accelerators, like the famous Higgs boson which has been predicted long time ago by theory \cite{higgs-mechanism1,higgs-mechanism2}, and has been confirmed by ATLAS and CMS collaborations in 2012 \cite{higgs-atlas, higgs-cms}. Despite its successes, there are still some open questions (Hierarchy problem, neutrinos mass, $\cdots$ etc) which are not yet answered by SM. To resolve these problems, many extensions of the SM were proposed. Most of these models predict new particles which are still not yet observed at experiments, as the vector-like quarks for example of which the single production at the Large Hadron Collider (LHC) is the main subject of this paper.

Vector-like quarks (VLQs) are spin-half fermions, where their left and right-handed components transform the same way under the electroweak symmetry group \cite{aguil1}. They are assumed to have the same colour quantum number as the ordinary quarks, which means that they behave exactly in the same way under the strong interaction. The vector-like nature of these particles allows to add their gauge invariant mass term without the need for the Higgs mechanism, i.e. their masses are not generated by the Yukawa couplings to the Higgs scalar doublet. The presence of vector-like-quarks is predicted by many models beyond the Standard Model (BSM) like extra-dimensions, little Higgs, composite Higgs and grand unification models \cite{extra-dim-vlqs, little-higgs-vlqs, composite-higgs-vlqs, grand-unif-vlqs}... etc. They are, mainly, introduced to treat the problem of naturalness of the electroweak scale and to stabilise the Higgs mass.
VLQs can be treated in model independent way at leading-order \cite{model-indep-vlqs-1, model-indep-vlqs-2, model-indep-vlqs-3, model-indep-vlqs-4, model-indep-vlqs-5}, where they are assumed to interact with SM quarks via the Yukawa mixing. This approach is based on an effective Lagrangian with a minimal set of free parameters, which enables us to study their phenomenology independently of the model \cite{model-indep-vlqs-4}. Extension to NLO QCD corrections of the model independent parameterisation was introduced in ref. \cite{model-indep-vlqs-6}. 

These new heavy states can be produced in different manners at the LHC. They can be produced in pairs, which are dominated by the QCD contribution and they are model independent. They can be produced singly in association with top quarks, jets, heavy gauge bosons or the Higgs. In contrast with the fourth generation of quarks, VLQs are consistent with the existing Higgs data and the recent LHC measurements. Due to their large mass and the fact that they can be produced in pairs via the strong interaction, which means relatively large cross sections, the research for such heavy states at the LHC is relatively accessible. 
One of the distinguishing characteristics of VLQs is the existence of flavour-changing neutral current, which open new particle production mechanisms and leads to widely diverse final states, which makes the investigation of these particles at the LHC very promising.
The search for VLQs has been analysed in many publications by ATLAS and CMS collaborations for LHC {\tt runI} and {\tt runII} \cite{THt1, THt2, TZt1, TZt2, TWb1, TWb2}. The analysis of recent data put strong bounds on the masses of these particles, the lower limit on the mass of the top quark partner, for example, is set between $715$-$950\, \text{GeV}$ \cite{bounds-vlqs}. 

In this paper we are interested in the production of a single top quark partner in association with a top quark in proton-proton collisions at leading and next-to-leading orders in the five and the six-flavour schemes (5FS and 6FS), and the decay of the top partner into a Higgs boson and a top quark, which leads to the interesting final state $t\bar t H$. We mention that the analysis of such final state, in SM and beyond, has received more and more attention by both theorists \cite{ref43,ref44} and experimentalists \cite{ref31,ref32,ref33,ref35,ref37} in the last few years, especially after the discovery of the Higgs boson. In section 2, we provide a brief overview of the model independent parameterisation and its extension to include QCD one loop corrections. In section 3, we calculate the total cross section and the differential distributions of the singly produced top partner in association with top quark in the 5FS. In section 4, we study the same signal but in the 6FS at the LHC and at the future hadron colliders ($\sqrt{s}=100\, \text{TeV}$). We discuss the implications of including top quark initiated processes at LO and NLO accuracies. In section 5, we analyse the decay product of the new heavy state to a Higgs boson and a top quark.
\section{VLQs model independent parameterisation overview}
\label{sec1}
\noindent
It is very important to have a model independent parameterisation which allows to study the main features and characteristics of VLQs for large variety of models. Actually, this was provided in \cite{model-indep-vlqs-4}, where the vector-like quarks are allowed to mix and decay to SM quarks via heavy weak gauge bosons and the Higgs. The main point of this approach, is that the interaction of such new heavy states with ordinary quarks is involved through the Yukawa couplings to the Higgs field doublet. In this strategy, the physics of VLQs is traced by an ensemble of few independent parameters handling the single production of these particles and their decay into heavy bosons and ordinary quarks. This parameterisation was extended to include QCD one loop corrections in \cite{model-indep-vlqs-6}, on which our present work is based. 

In this work, we are interested in the single production of the top partner $T$ in association with the top quark at the LHC, and the investigation of the associated Higgs with top quark pair final state which arises from the decay of the $T$ into a Higgs boson and a top quark (i.e. $pp\rightarrow T\bar{t}+\bar{T}t\rightarrow t\bar{t}H$). We consider many benchmark scenarios which might lead to such final states, where the $T$ is supposed to interact with the first or the third generations quarks. The general form of the Lagrangian, that we have to add to SM to describe those scenarios, can be expressed as follows:
\begin{align}
\call_{\scriptscriptstyle T}&=
-H[\bar T(\khl P_L+\khr P_R)q_u+\text{h.c.}]+\frac{g}{2c_{\scriptscriptstyle W}}[\bar T\not{\!\! Z}(\kzl P_L+\kzr P_R)q_u+\text{h.c.}]\nonumber\\
&+\frac{g}{\sqrt{2}}[\bar T\not{\!\! W}(\kwl P_L+\kwr P_R)q_d+\text{h.c.}]+i\bar T\not{\!\!D}T-m_{\scriptscriptstyle T} \bar TT.
\label{lagr}
\end{align}
where $q_u$ and $q_d$ are, respectively, up and down quark fields for a given generation. $D_{\mu}$ is the covariant derivative, $m_{\scriptscriptstyle T}$ is the mass of the top partner, $g$ is the weak coupling constant, $c_{\scriptscriptstyle W}$ is the cosine of the Weinberg mixing angle and $P_{L/R}$ are the left and right-handed chirality projectors.  
 The free parameters $\khlr$, $\kzlr$ and $\kwlr$ characterise the strength of the mixing between the $T$ and the quarks via the Higgs, the $Z$ and the $W$ bosons, respectively. In term of the branching fractions to SM quarks and heavy bosons, they are expressed as follows (the original parameterisation is given in \cite{model-indep-vlqs-4}):
 {\small
 \begin{align}
\khl&=\frac{\kappa_{\scriptscriptstyle T}m_{\scriptscriptstyle T}}{v}\sqrt{\frac{\zeta_i\xi_H^T}{\Gamma_H^T}}, & \khr&=\frac{\kappa_{\scriptscriptstyle T}m_t}{v}\sqrt{\frac{\zeta_i\xi_H^T}{\Gamma_H^T}},&
\kwl&=\kappa_{\scriptscriptstyle T}\sqrt{\frac{\zeta_i\xi_W^T}{\Gamma_W^T}}, & \kzl&=\kappa_{\scriptscriptstyle T}\sqrt{\frac{\zeta_i\xi_Z^T}{\Gamma_Z^T}}.\nonumber\\
 \xi_V^T&=\Gamma_V^T/(\sum_{\scriptscriptstyle V^{\prime}=Z, W, H}\Gamma_{V^{\prime}}^T),& \Gamma_{H}^T&\approx 1/2, & \Gamma_W^T&\approx 1, & \Gamma_Z^T&\approx 1/2.
\label{kappas}
 \end{align}
}
where $v$ and $m_t$ are, respectively, the Higgs vacuum expectation value and the top quark mass. The parameter $\zeta_i$ is the branching ratio into the $i^{\text{th}}$ SM quark generation. The parameter $\xi_V^T$ is the branching ratio into a $V$-boson (for $V=Z, W, H$). The quantities $\Gamma_V^T$ are the kinematic functions of the branching ratio of $T$ into a $V$-boson assuming that the quarks are massless. We mention that $\kwr$ and $\kzr$ are assumed to be negligible in this approach, see \cite{model-indep-vlqs-4} for more detail. The parameter $\kappa_{\scriptscriptstyle T}$ encodes information relative to the coupling strength to different generations. For mixing with quarks of the first and the third generations, we fix its value to $0.07$ and $0.1$, respectively, as suggested in ref. \cite{model-indep-vlqs-6}. For all the Standard Model quarks except the top quark, one can choose only one specific chirality since the mixing angle of the other one is suppressed by factor of $m_q/m_T$. However for the channel $T$-$H$-$t$, this is not a good approximation due to the top large mass, especially for top partner masses below $1\, \text{TeV}$. So one has to fix the parameters describing the mixing as in eq. (\ref{kappas}) assuming that the coupling of the sub-leading vertex is proportional to the same flavour mixing angle.

Finally, the Feynman rules, the ultraviolet counterterms, the reduction rational terms and the other required ingredients to compute one-loop QCD radiative corrections are generated automatically by the packages {\tt FeynRules}, {\tt NLOCT} and {\tt FeynArts} \cite{ref231, ref232, ref4, ref5} from the Lagrangian of the SM and the Lagrangian (\ref{lagr}). In this work, we generate a vector-like quark NLO model, where we include both right and left-handed couplings of $T$-$H$-$t$, as shown in eq. (\ref{kappas}), to take into account the effect of the top quark mass on the mixing of the top partner with the top quark via the Higgs boson. Furthermore, we employ the public UFO VLQ NLO model to verify our results (UFO for {\it Universal FeynRules Output})\footnote{The UFO and {\tt FeynRules} models are available in \url{http://feynrules.irmp.ucl.ac.be/wiki/NLOModels}.}. For more detail about the renormalization of the model, see ref. \cite{model-indep-vlqs-6}\footnote{In this work, we suppose that top quark partner interacts with SM quarks via the exchange of one, two or three heavy bosons, where the right-handed coupling of $T$-$H$-$t$ is included. The reactions $pp\rightarrow T\bar{t}+\bar{T}t$ and $pp\rightarrow T\bar{t}+\bar{T}t\rightarrow t\bar{t}H$ being examined for the first time in this model at NLO order in 5FS and 6FS.}.

\section{Top partner production in association with top quark at the LHC}
\label{sec3}
\noindent
\begin{figure}[tbp]
\centering
\includegraphics[width=10cm,height=3.5cm]{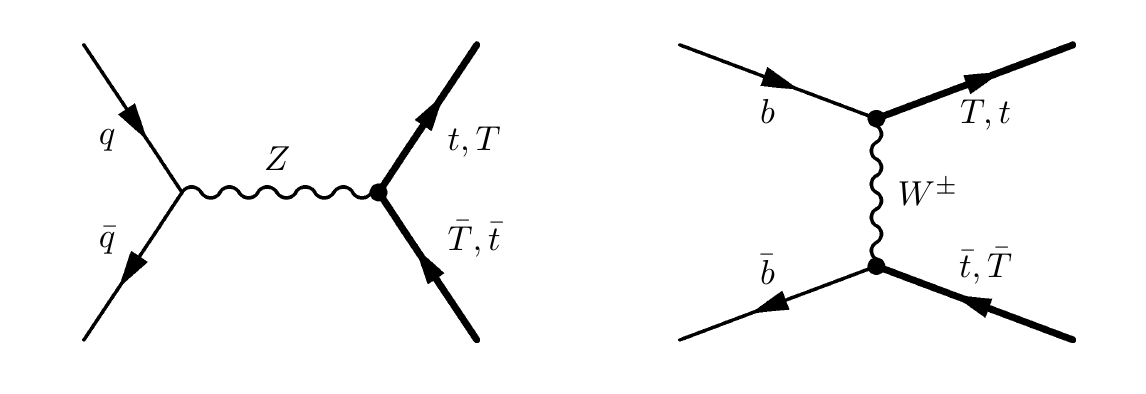}
    \caption{\footnotesize Leading-order Feynman diagrams for the single $T$ production in association with the top quark at proton-proton collisions: $q\bar q\rightarrow T\bar t+\bar T t$, where $q$ ($\bar q$) runs over all the massless quarks (anti-quarks).
    }
    \label{born}
   \end{figure}
\begin{figure}[tbp]
\centering
%
\includegraphics[width=11cm,height=1.5cm]{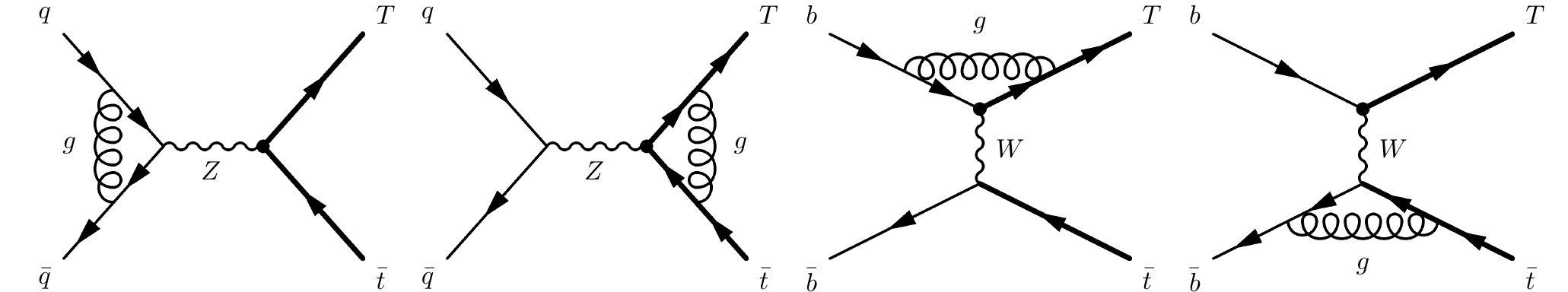}
\includegraphics[width=11cm,height=1.5cm]{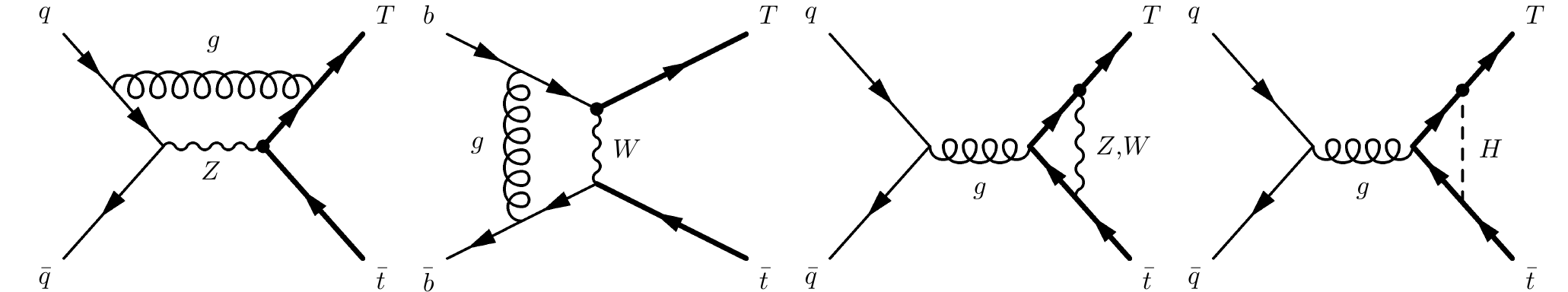}
\includegraphics[width=11cm,height=1.5cm]{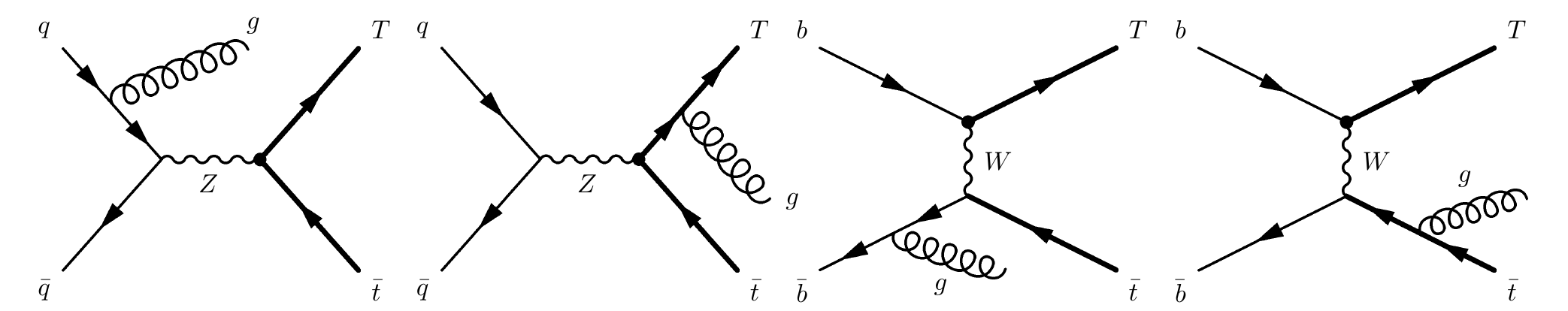}
\includegraphics[width=11cm,height=1.5cm]{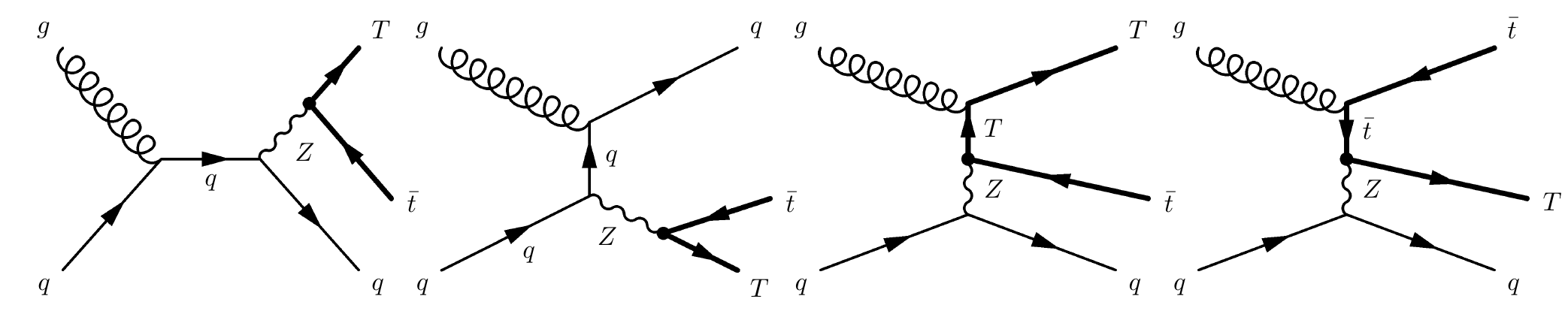}
\includegraphics[width=11cm,height=1.5cm]{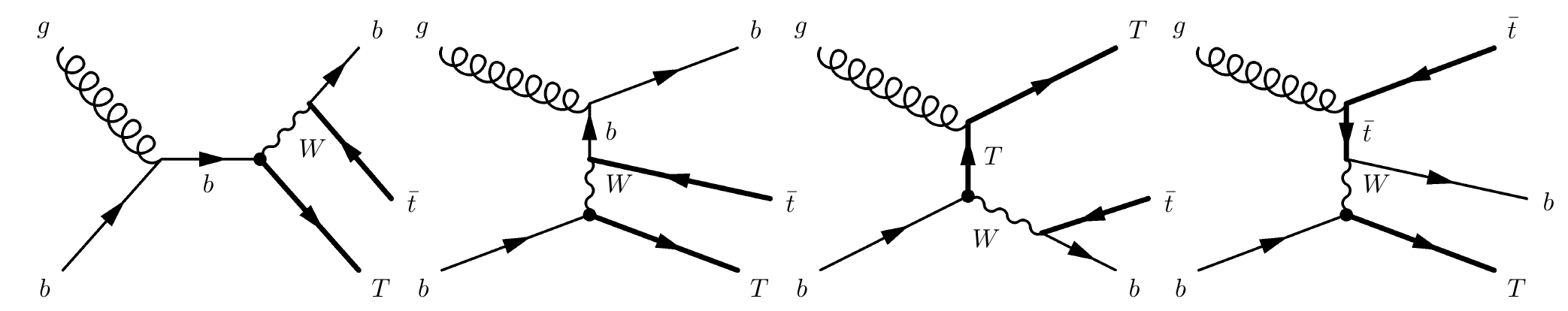}
\caption{\footnotesize Some of one-loop and real emission Feynman diagrams associated to both Born diagrams.}\label{loopreals}
%
\end{figure}

 
In this section, we compute the inclusive cross section and the differential distributions of the weak production of a single top partner in association with a top quark, at leading and next-to-leading orders in proton-proton collisions. 
The full NLO QCD cross section is calculated by interfacing the one-loop QCD corrections and the Born amplitudes after adding all the required UV counterterms and combining them with the squared amplitude of the real emission in the FKS subtraction framework \cite{fks1}. This can be done automatically by {\tt MadGraph} once the UFO VLQ NLO model is implemented, thanks to the automatic computation tools: {\tt FeynRules}, {\tt NLOCT}, {\tt FeynArts}, {\tt MadLoop}, {\tt MadFKS} and {\tt MadGraph} \cite{ref231, ref232, ref4, ref5, madloop, fks2, madgraph}. 
Let's call $\bf T^{\scriptscriptstyle\{0, 0, 3\}}_{\scriptscriptstyle\{Z, W, H\}}$ the benchmark scenario on which the vector like quark $T$ is allowed to mix with the third quark generations via the gauge weak bosons $Z, W$ and the Higgs. The lowest order processes, at $\mathcal{O}(\alpha^2)$, that lead to a final state with a single top quark partner associated with a top quark are,
\begin{align}
q\bar q&\rightarrow T\bar t, & q\bar q&\rightarrow \bar T t.
\end{align}
 In this scenario, the leading order weak production mechanism occurs by the exchange of $Z$ and $W$ bosons in the $s$-channel and the $t$-channel, respectively. We mention that the Higgs exchange in the $s$-channel is absent, since all the initial state quark flavours are massless. We point out that we work with the 5FS, where all the SM quarks are taken to be massless except the top. On top of that, we neglect all mixing between different generations of SM quarks since they give negligible contribution\footnote{We have checked this numerically at the LO order (hadronic cross section) by employing a UFO VLQ model, where all the CKM mixing matrix elements of the SM quarks have been implemented.}.  
The tree level Feynman diagrams describing those processes are shown in figure \ref{born}.  
The one-loop virtual amplitude, of order $\mathcal{O}(\alpha^2\alpha_s)$, is obtained by interfacing the weak Born Feynman diagrams, shown in figure \ref{born}, with all their one-loop QCD virtual correction diagrams with at least one virtual gluon running on the loop, i.e. we keep all the triangles and the boxes with at least two strong vertices. In the first and the second lines of figure \ref{loopreals}, we depict some of the one-loop Feynman diagrams. To suppress the ultraviolet divergences of this amplitude, one has to add all the necessary UV counterterms. This can be done automatically once the vector-like quarks renormalised model is implemented in {\tt MadGraph}. We notice that the interference of the Born Feynman diagrams with some of the one-loop Feynman diagrams gives vanishing contribution. For example, all the boxes interfaced with the Born diagram of the same channel are colour suppressed\footnote{The boxes must be included especially for $\bf T^{\scriptscriptstyle\{0, 0, 3\}}_{\scriptscriptstyle\{Z, W, 0\}}$ and $\bf T^{\scriptscriptstyle\{0, 0, 3\}}_{\scriptscriptstyle\{Z, W, H\}}$ to insure IR divergence cancellation.}, see the first two diagrams on the second line of figure \ref{loopreals}. This is due to the fact that the interference of the tree level Feynman diagram (without any strong vertex) and the box (with one internal virtual gluon) is proportional to two closed fermion loops with one strong vertex for each, which means that each one is proportional to $Tr(T^a)=0$ ($T^a$ are the generators of $SU(3)$ in the fundamental representation). Regarding the triangle loop diagrams without any virtual gluon running on the loop (see the last two diagrams on the second line of figure \ref{loopreals}), their interference with the $s$-channel Born Feynman diagram is colour suppressed for the same reasons as exposed above. Nevertheless, the interference with the $t$-channel Born Feynman diagram is not vanishing. In any case, such type of diagrams are not included in this calculation, since the VLQs NLO model that we employ does not include the necessary counterterms to perform mixed NLO QCD and electroweak corrections.

The interference of the Born and the virtual diagrams contains infrared (soft and/or collinear) divergent terms. For consistency of perturbation theory, one has to include diagrams with real emission of an extra parton, of order $\mathcal{O}(\alpha^2\alpha_s)$, at the partonic cross section level. The real emission processes are listed here,
\begin{align}
q\bar{q}&\rightarrow (T\bar t+\bar T t) g, & qg&\rightarrow (T\bar t+\bar T t) q, &\bar qg&\rightarrow (T\bar t+\bar T t) \bar q. 
\label{real}
\end{align}
Some of their corresponding Feynman diagrams are depicted in the last three lines of figure \ref{loopreals}. The first process in eq. (\ref{real}) is obtained from the Born diagrams by emitting an extra gluon by the initial or the final state quarks. The initial state emission can have soft or collinear singularities since the emitted and the emitter partons are both massless. However, the final state emission can only have soft singularities due to the non-vanishing mass of the heavy quarks ($T$ and $t$). After integrating the squared amplitude of these diagrams over the gluon momenta, we can show that it has exactly the same soft divergent terms, with opposite sign, as the one-loop virtual contribution interfaced with the Born. So, once these two contributions are combined together, the soft divergences are suppressed, thanks to Kinoshita-Lee-Nauenberg theorem (KLN) \cite{kln1, kln2}. Regarding the second and the third processes of eq. (\ref{real}), some of their Feynman diagrams are depicted in the last two lines of figure \ref{loopreals}. Only the second diagrams of the fourth and the fifth lines which are divergent (initial state collinear singularity). The remaining collinear singularities, after combining the virtual and real squared amplitudes, are factorized in the NLO parton distribution functions, thanks to the factorization theorem \cite{fact1, fact2}. Symbolically, the full NLO QCD cross section can be written as,
\begin{align}
\sigma^{\scriptscriptstyle\text{NLO}}=\int d\Phi_{2} d\sigma^{\scriptscriptstyle\text{Born}}&+\int d\Phi_{2} (d\sigma^{\scriptscriptstyle\text{virt}}+d\sigma^{\scriptscriptstyle\text{virt}}_{\scriptscriptstyle\text{sub}})+\int d\Phi_{2+1} (d\sigma^{\scriptscriptstyle\text{real}}+d\sigma^{\scriptscriptstyle\text{real}}_{\scriptscriptstyle\text{sub}})\nonumber\\
&+\int dx\int d\Phi_{2} (d\sigma^{\scriptscriptstyle\text{fact}}+d\sigma^{\scriptscriptstyle\text{fact}}_{\scriptscriptstyle\text{sub}}).
\label{sigmanlo}
\end{align}
where $d\Phi_{2}$ and $d\Phi_{2+1}$ are the two and three particle phase space of the final state; $\sigma^{\scriptscriptstyle\text{virt}}$ is the cross section of the virtual contribution (interference of the Born with the one loop diagrams), it is of order $\mathcal{O}(\alpha^2\alpha_s)$ and it is soft and collinear divergent; $\sigma^{\scriptscriptstyle\text{real}}$ is the cross section of the real emission, it is of the same order of $\sigma^{\scriptscriptstyle\text{virt}}$ and has the same soft divergent terms with opposite sign (squared amplitude of all the real emission diagrams) and it is also collinear divergent; $\sigma^{\scriptscriptstyle\text{fact}}$ is the collinear counterterms, it arises from the redefinition of the parton distribution functions. Each of these three contributions is IR divergent, however the sum of all of them is finite (factorization and KLN theorems). To be able to integrate numerically each contribution alone, we have to subtract the divergences from each one ($\sigma^{\scriptscriptstyle\text{virt, real, fact}}_{\scriptscriptstyle\text{sub}}$) so that they give no contribution to the full cross section as shown here, 
\begin{align}
\int d\Phi_{2} d\sigma^{\scriptscriptstyle\text{virt}}_{\scriptscriptstyle\text{sub}}+\int d\Phi_{2+1} d\sigma^{\scriptscriptstyle\text{real}}_{\scriptscriptstyle\text{sub}}+\int dx\int d\Phi_{2}d\sigma^{\scriptscriptstyle\text{fact}}_{\scriptscriptstyle\text{sub}}&=0.
\label{sub}
\end{align}
This can be done by employing a given subtraction method. In this paper, we use the FKS subtraction method which is automated by the module {\tt MadFKS} \cite{fks2}. Finally, the total inclusive cross section can be written as
\begin{align}
\sigma(T\bar t+\bar Tt)
&=\sigma_{\scriptscriptstyle Z}^{\scriptscriptstyle T\bar t+\bar Tt}+\sigma_{\scriptscriptstyle W}^{\scriptscriptstyle T\bar t+\bar Tt}+\sigma_{\scriptscriptstyle ZW}^{\scriptscriptstyle T\bar t+\bar Tt}.
\end{align}
where 
 $\sigma_{\scriptscriptstyle Z}^{\scriptscriptstyle T\bar t+\bar Tt}$ and $\sigma_{\scriptscriptstyle W}^{\scriptscriptstyle T\bar t+\bar Tt}$ are the inclusive (LO or NLO) hadronic cross sections for the singly produced top quark partner with the top through the exchange of $s$-channel $Z$ and $t$-channel $W$ bosons, respectively. $\sigma_{\scriptscriptstyle ZW}^{\scriptscriptstyle T\bar t+\bar Tt}$ is the interference contribution of the $s$ and $t$-channels Feynman diagrams. It contributes only for the scenarios $\bf T^{\scriptscriptstyle\{0, 0, 3\}}_{\scriptscriptstyle\{Z, W, 0\}}$ and $\bf T^{\scriptscriptstyle\{0, 0, 3\}}_{\scriptscriptstyle\{Z, W, H\}}$ and for processes initiated by $b$-quarks. For the later one, the inclusion of the boxes is indispensable to insure IR divergence cancellation. 

\begin{figure}[tbp]
\centering
\includegraphics[width=13cm,height=6cm]{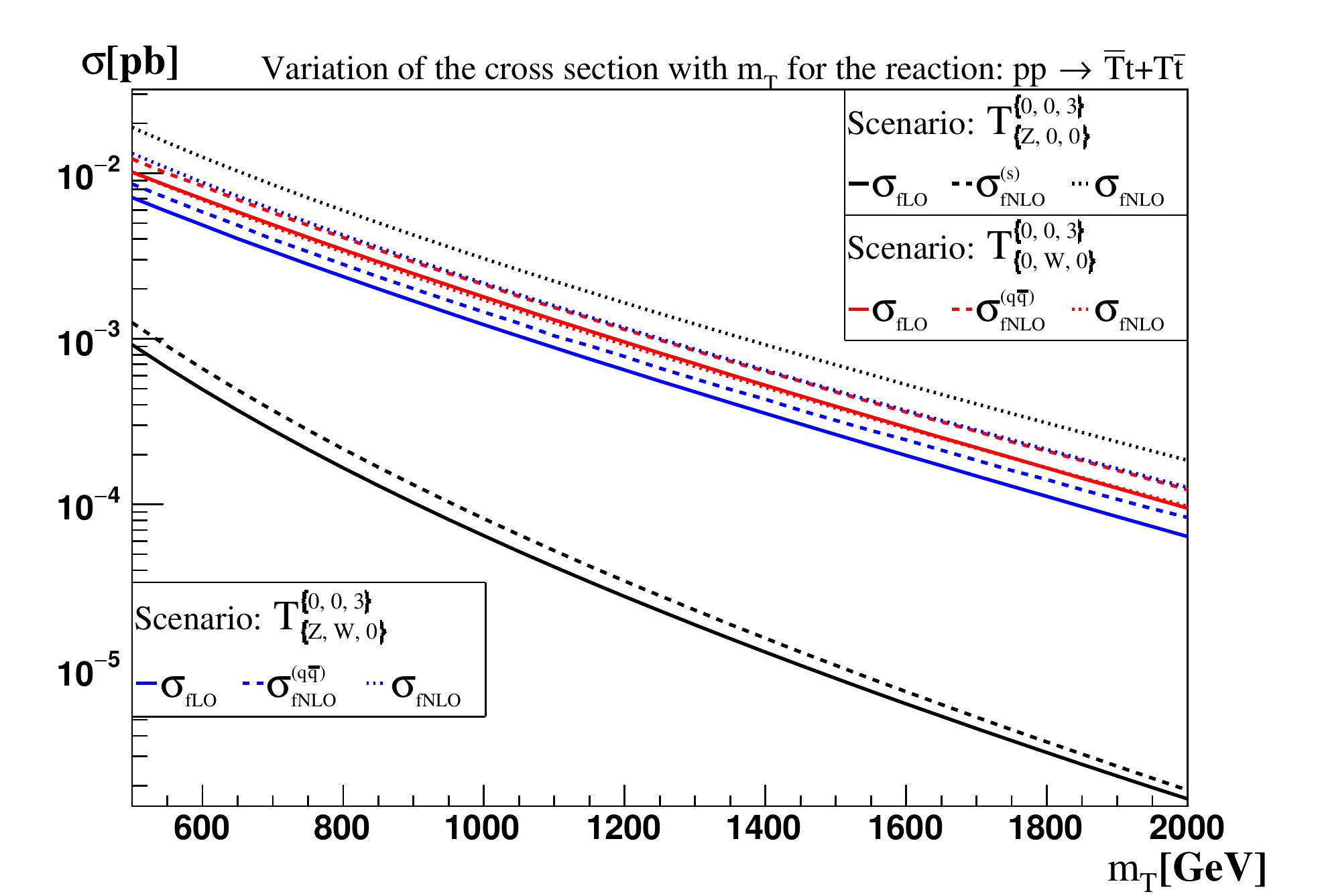}
\caption{\footnotesize Variation of the fLO and fNLO cross sections in term of the top partner mass for the three benchmark scenarios: $\bf T^{\scriptscriptstyle\{0, 0, 3\}}_{\scriptscriptstyle\{Z, 0, 0\}}$, $\bf T^{\scriptscriptstyle\{0, 0, 3\}}_{\scriptscriptstyle\{0, W, 0\}}$ and $\bf T^{\scriptscriptstyle\{0, 0, 3\}}_{\scriptscriptstyle\{Z, W, 0\}}$ (at center-of-mass energy $\sqrt{s}=13\, \text{TeV}$).}
\label{sigma-mass}
\end{figure}
\begin{figure}[tbp]
\centering
\includegraphics[width=7.75cm,height=5cm]{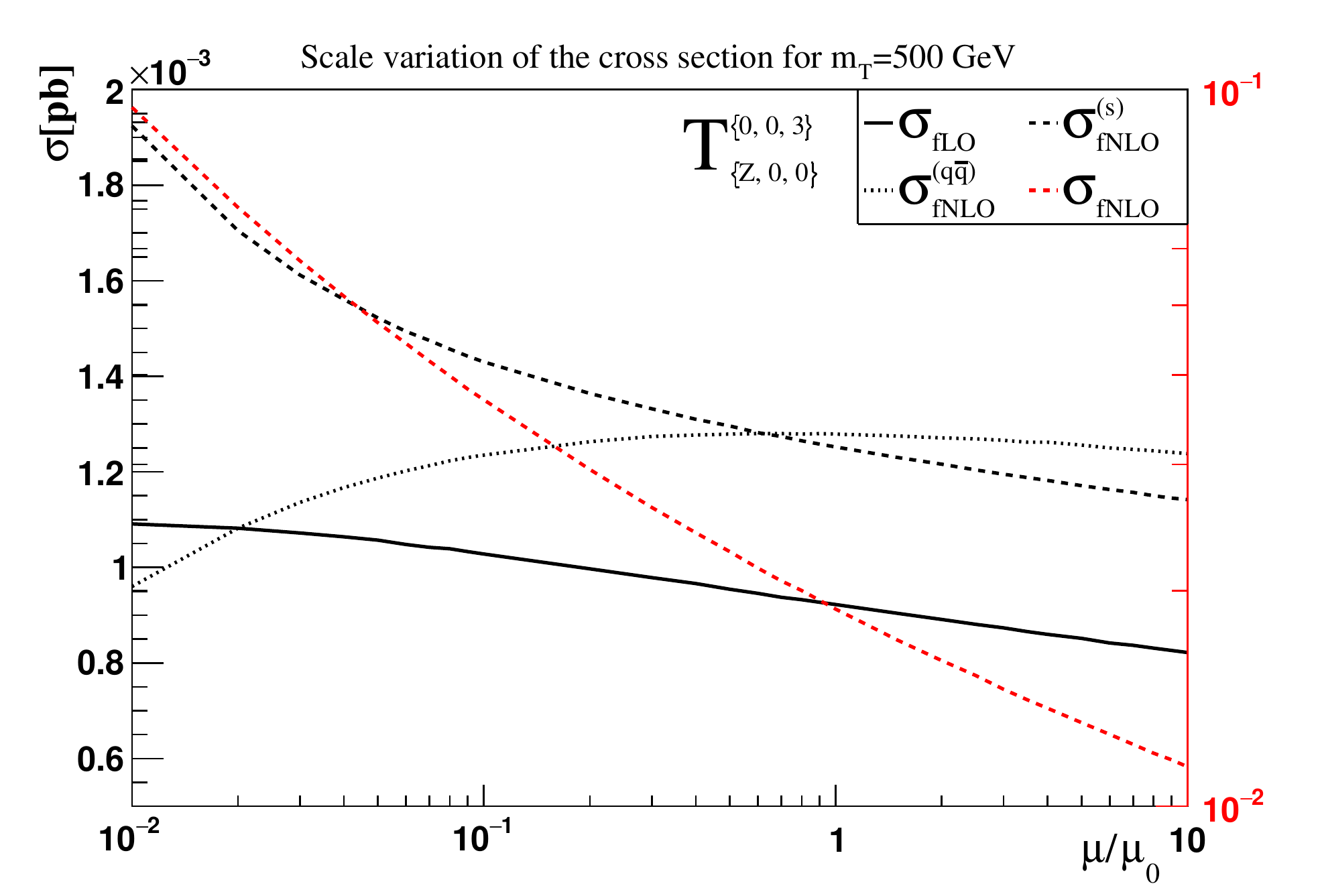}\includegraphics[width=7.75cm,height=5cm]{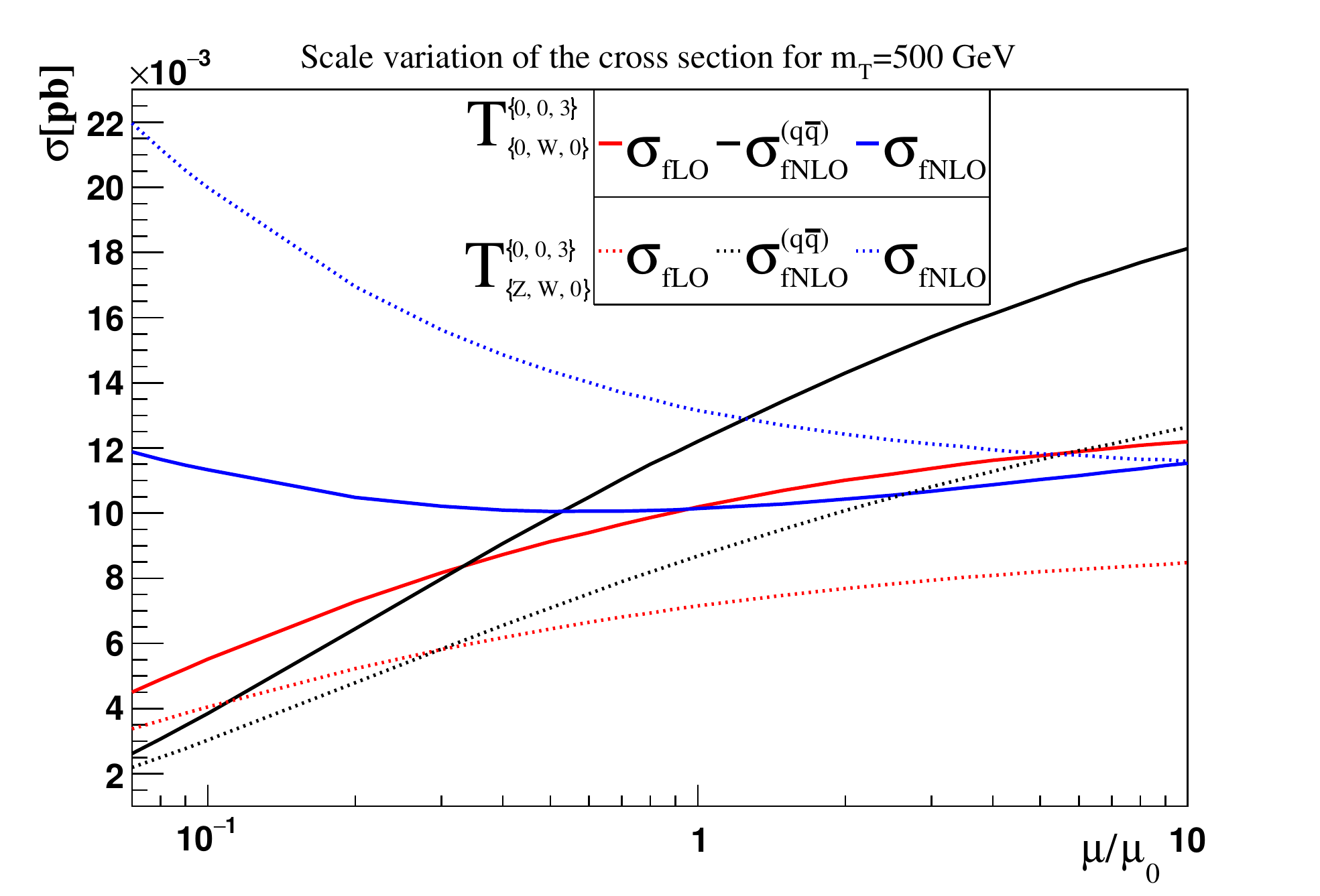}
\caption{\footnotesize Scale variation of the total cross section for the benchmark scenarios: $\bf T^{\scriptscriptstyle\{0, 0, 3\}}_{\scriptscriptstyle\{Z, 0, 0\}}$, $\bf T^{\scriptscriptstyle\{0, 0, 3\}}_{\scriptscriptstyle\{0, W, 0\}}$ and $\bf T^{\scriptscriptstyle\{0, 0, 3\}}_{\scriptscriptstyle\{Z, W, 0\}}$ ($m_{\scriptscriptstyle T}=500\, \text{GeV}$). We notice that the red-dashed plot on the left-side follows the Y-axis on the right-side of the panel.}
\label{sigma-mu}
\end{figure}

\begin{table*}[tbp]
\centering
 \renewcommand{\arraystretch}{1.40}
 \setlength{\tabcolsep}{10pt}
 \begin{adjustbox}{width=0.97\textwidth}
 \footnotesize
 \begin{tabular}{!{\vrule width 1pt}ll!{\vrule width 1pt}lllll!{\vrule width 1pt}}
  \noalign{\hrule height 1pt}
    \bf $m_{\scriptscriptstyle T}$ [GeV] &\bf Scenario &\bf $\sigma_{\scriptscriptstyle\text{fLO}}$ [pb] &\bf $\sigma_{\scriptscriptstyle\text{fNLO}}$ [pb]&\bf$K$-factor&\bf$\sigma_{\scriptscriptstyle\text{fNLO}}^{\scriptscriptstyle\text{(s)}}$ [pb]&\bf$K$-factor\\
  \noalign{\hrule height 1pt}
   $500$ & {\footnotesize\bf $\bf T^{\scriptscriptstyle\{0, 0, 3\}}_{\scriptscriptstyle\{Z, 0, 0\}}$} & $(9.230\times 10^{-4}){}^{+3.5\%}_{-3.4\%}{}^{+6.3\%}_{-6.3\%}$ & $(1.894\times 10^{-2}){}^{+19.8\%}_{-15.3\%}{}^{+1.5\%}_{-1.5\%}$&20.52&$(1.252\times 10^{-3}){}^{+3.4\%}_{-3.0\%}{}^{+1.9\%}_{-1.9\%}$&1.36\\
   \hdashline
   $1000$ & {\footnotesize\bf $\bf T^{\scriptscriptstyle\{0, 0, 3\}}_{\scriptscriptstyle\{Z, 0, 0\}}$} &  $(6.456\times 10^{-5}){}^{+6.9\%}_{-6.2\%}{}^{+6.7\%}_{-6.7\%}$ &  $(3.053\times 10^{-3}){}^{+23.3\%}_{-17.6\%}{}^{+2.8\%}_{-2.8\%}$&47.29&$(8.201\times10^{-5}){}^{+4.1\%}_{-3.9\%}{}^{+2.5\%}_{-2.5\%}$&1.27\\
   \hdashline
   $1500$ & {\footnotesize\bf $\bf T^{\scriptscriptstyle\{0, 0, 3\}}_{\scriptscriptstyle\{Z, 0, 0\}}$} & $(8.877\times 10^{-6}){}^{+9.2\%}_{-8.0\%}{}^{+7.9\%}_{-7.9\%}$  & $(6.948\times 10^{-4}){}^{+25.5\%}_{-19.0\%}{}^{+4.4\%}_{-4.4\%}$&78.27&$(1.065\times10^{-5}){}^{+4.7\%}_{-4.6\%}{}^{+3.5\%}_{-3.5\%}$&1.20\\
   \hdashline
   $2000$ & {\footnotesize\bf $\bf T^{\scriptscriptstyle\{0, 0, 3\}}_{\scriptscriptstyle\{Z, 0, 0\}}$} & $(1.663\times 10^{-6}){}^{+11.0\%}_{-9.4\%}{}^{+9.8\%}_{-9.8\%}$& $(1.846\times 10^{-4}){}^{+27.3\%}_{-20.0\%}{}^{+6.3\%}_{-6.3\%}$&111.00&$(1.880\times10^{-6}){}^{+5.2\%}_{-5.1\%}{}^{+5.1\%}_{-5.1\%}$&1.13\\
\hline
   $500$ & {\footnotesize\bf $\bf T^{\scriptscriptstyle\{0, 0, 3\}}_{\scriptscriptstyle\{Z, 0, H\}}$} & $(4.614\times 10^{-4}){}^{+3.5\%}_{-3.4\%}{}^{+6.3\%}_{-6.3\%}$ & $(9.452\times 10^{-3}){}^{+19.8\%}_{-15.4\%}{}^{+1.5\%}_{-1.5\%}$&20.49&$(6.255\times10^{-4}){}^{+3.4\%}_{-3.0\%}{}^{+1.9\%}_{-1.9\%}$&1.36\\
   \hdashline
   $1000$ & {\footnotesize\bf $\bf T^{\scriptscriptstyle\{0, 0, 3\}}_{\scriptscriptstyle\{Z, 0, H\}}$} & $(3.230\times 10^{-5}){}^{+6.9\%}_{-6.2\%}{}^{+6.7\%}_{-6.7\%}$ & $(1.527\times 10^{-3}){}^{+23.3\%}_{-17.6\%}{}^{+2.8\%}_{-2.8\%}$&47.28&$(4.106\times10^{-5}){}^{+4.1\%}_{-3.9\%}{}^{+2.5\%}_{-2.5\%}$&1.27\\
   \hdashline
   $1500$ & {\footnotesize\bf $\bf T^{\scriptscriptstyle\{0, 0, 3\}}_{\scriptscriptstyle\{Z, 0, H\}}$} & $(4.438\times 10^{-6}){}^{+9.2\%}_{-8.0\%}{}^{+7.9\%}_{-7.9\%}$  & $(3.475\times 10^{-4}){}^{+25.5\%}_{-18.9\%}{}^{+4.4\%}_{-4.4\%}$&78.30&$(5.327\times10^{-6}){}^{+4.6\%}_{-4.5\%}{}^{+3.4\%}_{-3.4\%}$&1.20\\
   \hdashline
   $2000$ & {\footnotesize\bf $\bf T^{\scriptscriptstyle\{0, 0, 3\}}_{\scriptscriptstyle\{Z, 0, H\}}$} & $(8.316\times 10^{-7}){}^{+11.0\%}_{-9.4\%}{}^{+9.8\%}_{-9.8\%}$& $(9.225\times 10^{-5}){}^{+27.2\%}_{-20.0\%}{}^{+6.3\%}_{-6.3\%}$&111.93&$(9.386\times10^{-7}){}^{+5.2\%}_{-5.2\%}{}^{+5.2\%}_{-5.2\%}$&1.13\\
\noalign{\hrule height 1pt}
   \bf$m_{\scriptscriptstyle T}$~[GeV] &\bf Scenario &\bf $\sigma_{\scriptscriptstyle\text{fLO}}$ [pb] &\bf $\sigma_{\scriptscriptstyle\text{fNLO}}$ [pb]&\bf$K$-factor&\bf$\sigma_{\scriptscriptstyle\text{fNLO}}^{\scriptscriptstyle (q\bar{q})}$ [pb]&\bf$K$-factor\\
 \noalign{\hrule height 1pt}
   $500$ & {\footnotesize\bf $\bf T^{\scriptscriptstyle\{0, 0, 3\}}_{\scriptscriptstyle\{0, W, 0\}}$} & $( 1.018\times 10^{-2}){}^{+8.0\%}_{-10.4\%}{}^{+21.9\%}_{-21.9\%}$ & $(1.013\times 10^{-2}){}^{+3.0\%}_{-1.4\%}{}^{+5.0\%}_{-5.0\%}$&0.99&$(1.222\times 10^{-2}){}^{+20.2\%}_{-19.3\%}{}^{+4.6\%}_{-4.6\%}$&1.20\\
   \hdashline
   $1000$ & {\footnotesize\bf $\bf T^{\scriptscriptstyle\{0, 0, 3\}}_{\scriptscriptstyle\{0, W, 0\}}$} & $(1.788\times 10^{-3}){}^{+3.1\%}_{-5.0\%}{}^{+32.2\%}_{-32.2\%}$ &$(1.716\times 10^{-3}){}^{+3.0\%}_{-1.3\%}{}^{+9.3\%}_{-9.3\%}$&0.96&$(2.121\times 10^{-3}){}^{+16.2\%}_{-15.7\%}{}^{+8.6\%}_{-8.6\%}$&1.19\\
   \hdashline
   $1500$ & {\footnotesize\bf $\bf T^{\scriptscriptstyle\{0, 0, 3\}}_{\scriptscriptstyle\{0, W, 0\}}$} & $(3.907\times 10^{-4}){}^{+0.0\%}_{-1.6\%}{}^{+44.5\%}_{-44.5\%}$ & $(3.813\times 10^{-4}){}^{+2.9\%}_{-1.7\%}{}^{+14.7\%}_{-14.7\%}$&0.98&$(4.768\times 10^{-4}){}^{+13.9\%}_{-13.5\%}{}^{+13.7\%}_{-13.7\%}$&1.22\\
   \hdashline
   $2000$ & {\footnotesize\bf $\bf T^{\scriptscriptstyle\{0, 0, 3\}}_{\scriptscriptstyle\{0, W, 0\}}$} & $(9.474\times 10^{-5}){}^{+1.1\%}_{-2.3\%}{}^{+58.3\%}_{-58.3\%}$& $(9.716\times 10^{-5}){}^{+2.7\%}_{-2.0\%}{}^{+21.6\%}_{-21.6\%}$&1.03&$(1.228\times 10^{-4}){}^{+12.3\%}_{-11.9\%}{}^{+20.2\%}_{-20.2\%}$&1.30\\
      \hline
   $500$ & {\footnotesize\bf $\bf T^{\scriptscriptstyle\{0, 0, 3\}}_{\scriptscriptstyle\{0, W, H\}}$} & $(6.786\times 10^{-3}){}^{+8.0\%}_{-10.4\%}{}^{+21.8\%}_{-21.8\%}$ & $(6.755\times 10^{-3}){}^{+3.0\%}_{-1.4\%}{}^{+5.0\%}_{-5.0\%}$&0.99&$(8.153\times 10^{-3}){}^{+20.1\%}_{-19.3\%}{}^{+4.6\%}_{-4.6\%}$&1.20\\
   \hdashline
   $1000$ & {\footnotesize\bf $\bf T^{\scriptscriptstyle\{0, 0, 3\}}_{\scriptscriptstyle\{0, W, H\}}$} &  $(1.191\times 10^{-3}){}^{+3.1\%}_{-5.0\%}{}^{+32.0\%}_{-32.0\%}$& $(1.147\times 10^{-3}){}^{+3.0\%}_{-1.3\%}{}^{+9.3\%}_{-9.3\%}$&0.96&$(1.410\times 10^{-3}){}^{+16.1\%}_{-15.7\%}{}^{+8.6\%}_{-8.6\%}$&1.18\\
   \hdashline
   $1500$ & {\footnotesize\bf $\bf T^{\scriptscriptstyle\{0, 0, 3\}}_{\scriptscriptstyle\{0, W, H\}}$} &  $(2.604\times 10^{-4}){}^{+0.0\%}_{-1.6\%}{}^{+44.4\%}_{-44.4\%}$ & $(2.554\times 10^{-4}){}^{+2.9\%}_{-1.7\%}{}^{+14.7\%}_{-14.7\%}$&0.98&$(3.179\times 10^{-4}){}^{+13.9\%}_{-13.5\%}{}^{+13.8\%}_{-13.8\%}$&1.22\\
   \hdashline
   $2000$ & {\footnotesize\bf $\bf T^{\scriptscriptstyle\{0, 0, 3\}}_{\scriptscriptstyle\{0, W, H\}}$} & $(6.329\times 10^{-5}){}^{+1.1\%}_{-2.3\%}{}^{+58.3\%}_{-58.3\%}$& $(6.476\times 10^{-5}){}^{+2.7\%}_{-2.0\%}{}^{+21.6\%}_{-21.6\%}$&1.02&$(8.188\times 10^{-5}){}^{+12.3\%}_{-11.9\%}{}^{+20.2\%}_{-20.2\%}$&1.29\\
\noalign{\hrule height 1pt}
   \bf$m_{\scriptscriptstyle T}$~[GeV] &\bf Scenario &\bf $\sigma_{\scriptscriptstyle\text{fLO}}$ [pb] &\bf $\sigma_{\scriptscriptstyle\text{fNLO}}$ [pb]&\bf$K$-factor&\bf$\sigma_{\scriptscriptstyle\text{fNLO}}^{\scriptscriptstyle\text{($q\bar{q}$)}}$ [pb]&\bf$K$-factor\\
 \noalign{\hrule height 1pt}
   $500$ & {\footnotesize\bf $\bf T^{\scriptscriptstyle\{0, 0, 3\}}_{\scriptscriptstyle\{Z, W, 0\}}$} & $(7.139\times 10^{-3}){}^{+7.5\%}_{-9.8\%}{}^{+21.0\%}_{-21.0\%}$ & $(1.319\times 10^{-2}){}^{+9.1\%}_{-5.8\%}{}^{+3.2\%}_{-3.2\%}$&1.85&$(8.661\times 10^{-3}){}^{+19.4\%}_{-18.2\%}{}^{+4.3\%}_{-4.3\%}$&1.21\\
   \hdashline
   $1000$ & {\footnotesize\bf $\bf T^{\scriptscriptstyle\{0, 0, 3\}}_{\scriptscriptstyle\{Z, W, 0\}}$} & $(1.218\times 10^{-3}){}^{+2.9\%}_{-4.8\%}{}^{+31.7\%}_{-31.7\%}$ & $(2.173\times 10^{-3}){}^{+10.3\%}_{-6.7\%}{}^{+6.1\%}_{-6.1\%}$&1.78&$(1.447\times 10^{-3}){}^{+15.9\%}_{-15.3\%}{}^{+8.4\%}_{-8.4\%}$&1.19\\
   \hdashline
   $1500$ & {\footnotesize\bf $\bf T^{\scriptscriptstyle\{0, 0, 3\}}_{\scriptscriptstyle\{Z, W, 0\}}$} & $(2.636\times 10^{-4}){}^{+0.0\%}_{-1.5\%}{}^{+44.0\%}_{-44.0\%}$ & $(4.875\times 10^{-4}){}^{+11.2\%}_{-7.5\%}{}^{+9.7\%}_{-9.7\%}$&1.85&$(3.219\times 10^{-4}){}^{+13.7\%}_{-13.3\%}{}^{+13.6\%}_{-13.6\%}$&1.22\\
   \hdashline
   $2000$ & {\footnotesize\bf $\bf T^{\scriptscriptstyle\{0, 0, 3\}}_{\scriptscriptstyle\{Z, W, 0\}}$} & $(6.373\times 10^{-5}){}^{+1.2\%}_{-2.3\%}{}^{+57.9\%}_{-57.9\%}$ & $(1.266\times 10^{-4}){}^{+12.2\%}_{-8.4\%}{}^{+14.0\%}_{-14.0\%}$&1.99&$(8.269\times 10^{-5}){}^{+12.2\%}_{-11.7\%}{}^{+20.0\%}_{-20.0\%}$&1.30\\
 \hline
   $500$ & {\footnotesize\bf $\bf T^{\scriptscriptstyle\{0, 0, 3\}}_{\scriptscriptstyle\{Z, W, H\}}$} & $(5.359\times 10^{-3}){}^{+7.5 \%}_{-9.8\%}{}^{+21.0\%}_{-21.0\%}$ & $(9.885\times 10^{-3}){}^{+9.1\%}_{-5.8\%}{}^{+3.2\%}_{-3.2\%}$&1.84&$(6.509\times10^{-3}){}^{+19.4 \%}_{-18.2\%}{}^{+4.3\%}_{-4.3\%}$&1.22\\
   \hdashline
   $1000$ & {\footnotesize\bf $\bf T^{\scriptscriptstyle\{0, 0, 3\}}_{\scriptscriptstyle\{Z, W, H\}}$} & $(9.132\times 10^{-4}){}^{+2.9\%}_{-4.8\%}{}^{+31.7\%}_{-31.7\%}$ &$(1.626\times 10^{-3}){}^{+10.3\%}_{-6.7\%}{}^{+6.1\%}_{-6.1\%}$&1.78&$(1.082\times 10^{-3}){}^{+15.9\%}_{-15.4\%}{}^{+8.4\%}_{-8.4\%}$&1.18\\
   \hdashline
   $1500$ & {\footnotesize\bf $\bf T^{\scriptscriptstyle\{0, 0, 3\}}_{\scriptscriptstyle\{Z, W, H\}}$} & $(1.978\times 10^{-4}){}^{+0.0\%}_{-1.5\%}{}^{+44.0\%}_{-44.0\%}$  & $(3.649\times 10^{-4}){}^{+11.3\%}_{-7.5\%}{}^{+9.6\%}_{-9.6\%}$&1.84&$(2.416\times 10^{-4}){}^{-13.8\%}_{+13.3\%}{}^{+13.5\%}_{-13.5\%}$&1.22\\
   \hdashline
   $2000$ & {\footnotesize\bf $\bf T^{\scriptscriptstyle\{0, 0, 3\}}_{\scriptscriptstyle\{Z, W, H\}}$} & $(4.787\times 10^{-5}){}^{+1.2\%}_{-2.3\%}{}^{+57.9\%}_{-57.9\%}$& $(9.493\times 10^{-5}){}^{+12.3\%}_{-8.4\%}{}^{+14.0\%}_{-14.0\%}$&1.98&$(6.199\times 10^{-5}){}^{+12.2\%}_{-11.8\%}{}^{+20.0\%}_{-20.0\%}$&1.29\\
 \noalign{\hrule height 1pt}
  \end{tabular}
  \end{adjustbox}
  \caption{\footnotesize fLO and fNLO QCD inclusive cross sections for
   $T\bar t+\bar T t$ production at $\sqrt{s}=13$ TeV. The results are shown together with their associated scale (left)
   and PDF (right) uncertainties. The values of the external parameters are fixed to: $m_H= 125\, \text{GeV}$, $m_Z=91.1876\, \text{GeV}$, $m_t= 173.1\, \text{GeV}$, $\alpha_s(m_Z)=0.118$ and $G_F=1.16637\times 10^{-5}\, \text{GeV}^{-2}$,
where $\alpha_s(m_Z)$ is the strong coupling constant at the $Z$ mass.}
   \label{scen16}
  \end{table*}
We use {\tt MadGraph5aMc@NLO} version 2.5.4, where the {\tt NNPDF 3.0} PDF set is adopted (we use {\tt NNPDF30\_lo\_as\_0118} for LO and {\tt NNPDF30\_nlo\_as\_0118} for NLO) \cite{lhapdf6, nnpdf}. The couplings of the top partner to the third generation quarks are defined in eq. (\ref{kappas}). 
We use the dynamical scale scheme, where we set the renormalization ($\mu_R$) and factorization ($\mu_F$) scales equal to each other ($\mu_R=\mu_F=\mu=M_T/2$, where $M_T$ is the transverse mass of the final state particles system). 
To estimate the theoretical uncertainties, we evaluate the cross section for the three values of the scale $\mu/2$, $\mu$ and $2\mu$.
In table \ref{scen16}, we furnish the leading and the next-to-leading fixed orders (fLO and fNLO) cross sections with their relative scale and PDF uncertainties, for the production of a top quark partner in association with a top quark for different values of $m_{\scriptscriptstyle T}$ in six benchmark scenarios\footnote{We mention that the lower limits of the $T$ mass range between $715$-$950\, \text{GeV}$ according to recent LHC data. Nevertheless, in this work, we vary it between $500\,\text{GeV}$ and $2000\, \text{GeV}$ to better study the behaviour of the cross section and to compare the scale uncertainties for low and high $T$ mass.}. We notice that $\sigma_{\scriptscriptstyle\text{fNLO}}$ is the full fNLO cross section, where all the real emission processes are taken into account. On the other hand, for $\sigma_{\scriptscriptstyle\text{fNLO}}^{\scriptscriptstyle\text{(s)}}$ the $t$-channel gluon-quark (anti-quark) real emission corrections to the $s$-channel Born Feynman diagrams are omitted, and for $\sigma_{\scriptscriptstyle\text{fNLO}}^{\scriptscriptstyle\text{($q\bar{q}$)}}$ all the gluon-quark (anti-quark) initiated processes are ignored. The first thing that we observe, from table \ref{scen16} and figure \ref{sigma-mass}, is the huge difference between $\sigma_{\scriptscriptstyle\text{fLO}}$ and $\sigma_{\scriptscriptstyle\text{fNLO}}$ for the scenarios $\bf T^{\scriptscriptstyle\{0, 0, 3\}}_{\scriptscriptstyle\{Z, 0, 0\}}$ and $\bf T^{\scriptscriptstyle\{0, 0, 3\}}_{\scriptscriptstyle\{Z, 0, H\}}$ which leads to gigantic $K$-factors (increasing with the $T$ mass). This is due to the fact that the $t$-channel gluon-quark (anti-quark) Feynman diagrams contribution (seen as a new Born contribution), which appears exclusively at NLO calculation, dominates over the $s$-channel tree level and one loop Feynman diagram contributions. If one omit the real emission $t$-channel contribution, as shown on the last two columns of the first part of table \ref{scen16} ($\sigma_{\scriptscriptstyle\text{fNLO}}^{\scriptscriptstyle\text{($s$)}}$), the $K$-factors for the different masses are reduced (1.36 for $m_{\scriptscriptstyle T}=500\,\text{GeV}$ and 1.13 for $m_{\scriptscriptstyle T}=2000\,\text{GeV}$). As a matter of fact, the squared amplitudes of the $t$-channel real emission Feynman diagrams, the last two graphs in the fourth line of figure \ref{loopreals}, are proportional to $m_Z^{-4}$ (for small $t$-channel $Z$ boson virtuality). On the other hand, the squared amplitudes of the  $s$-channel Feynman diagrams (neglecting the interference contribution) are proportional to $(\hat s-m_{Z}^2)^{-2}$, where $\hat s \geq (m_T+m_t)^2$. On top of that, the $t$-channel contribution comes from the gluon-quark initiated processes, which make the former contribution very dominant and almost control the behaviour of the so called fixed order NLO QCD cross section, see figure \ref{sigma-mass}. One of the major purposes of NLO calculation is the reduction of the renormalization and factorization scales dependency of the cross section. From this table and figure \ref{sigma-mu}, we see that the scale dependence is reduced only if we suppress the $t$-channel gluon-quark (anti-quark) real emission Feynman diagrams contribution for the scenarios $\bf T^{\scriptscriptstyle\{0, 0, 3\}}_{\scriptscriptstyle\{Z, 0, 0\}}$  and $\bf T^{\scriptscriptstyle\{0, 0, 3\}}_{\scriptscriptstyle\{Z, 0, H\}}$\footnote{If we suppress the contribution of gluon-quark (anti-quark) real emission processes, the scale dependency is much more improved as shown on the left side of figure \ref{sigma-mu} (dotted black curve). Nevertheless, this is not physical since these processes contribute to the scaling violation of the quark distribution functions.}. If we consider such contribution, the fNLO cross section behaves on the scale as LO-like cross section i.e. it is monotonically decreasing on the scale (see the red-dashed curve on the left side of figure \ref{sigma-mu}). So, to reduce the scale dependency, one has to calculate the 
higher order 
corrections (NNLO for example), which is not the purpose of this paper. Another alternative to reduce the $K$-factors is to restrict the phase space of the real emission contribution. For example, by enforcing the transverse momentum of the jet to be $P_T(j)>m_{\scriptscriptstyle T}^{\scriptscriptstyle(i)}/(3/2+i)$ for $i=1, \cdots 4$, the $K$-factors are reduced to $1.40$ for $m_{\scriptscriptstyle T}^{\scriptscriptstyle(1)}=500\, \text{GeV}$, $1.39$ for $m_{\scriptscriptstyle T}^{\scriptscriptstyle(2)}=1000\, \text{GeV}$, $1.56$ for $m_{\scriptscriptstyle T}^{\scriptscriptstyle(3)}=1500\, \text{GeV}$ and $1.70$ for $m_{\scriptscriptstyle T}^{\scriptscriptstyle(4)}=2000\, \text{GeV}$. In fact, these cuts restrict the virtuality of the $Z$ boson (especially $t$-channel exchange) to be much larger than its mass which makes the contribution of the real emission (especially gluon-quark initiated processes) sub-dominant. Regarding the scenarios 
$\bf T^{\scriptscriptstyle\{0, 0, 3\}}_{\scriptscriptstyle\{0, W, 0\}}$ and $\bf T^{\scriptscriptstyle\{0, 0, 3\}}_{\scriptscriptstyle\{0, W, H\}}$, all the partonic processes contributing to them are initiated by bottom quark pairs or gluon-bottom (anti-bottom). From the second part of this table, we see that the real and virtual contributions almost cancel together which leads to $K$-factors nearly equal to one. On the other hand, the scale dependency is improved especially for relatively low top partner mass for both scenarios (see the blue solid curve on the right side of figure \ref{sigma-mu}). In what concern the scenarios $\bf T^{\scriptscriptstyle\{0, 0, 3\}}_{\scriptscriptstyle\{Z, W, 0\}}$ and $\bf T^{\scriptscriptstyle\{0, 0, 3\}}_{\scriptscriptstyle\{Z, W, H\}}$, the full fNLO cross section is about the double of the fLO one. This is attributed to the quark-gluon (anti-quark) initiated processes contribution which arises only at NLO and it is known to increase the $K$-factor. If the later contribution is suppressed (see the last two columns of the third part of table \ref{scen16}, the $K$-factors are reduced to be of order 1.2. Concerning the uncertainties associated to the parton distribution functions, they increase with the mass of the top partner for both fLO and fNLO cross sections and for all the benchmark scenarios\footnote{For benchmark scenarios where the $W$ boson is present, the PDF uncertainties are very large compared to the other scenarios. This is due to the bottom quark PDF which shows the largest difference between different sets for large Bjorken-$x$ (especially the LO PDF).}. This is due to the fact that, for large Bjorken-$x$, the errors of the different PDF sets become larger as a result of the lack of data in this region.

\begin{figure}[tbp]
 \centering
 {
\includegraphics[width=7.5cm,height=5.5cm]{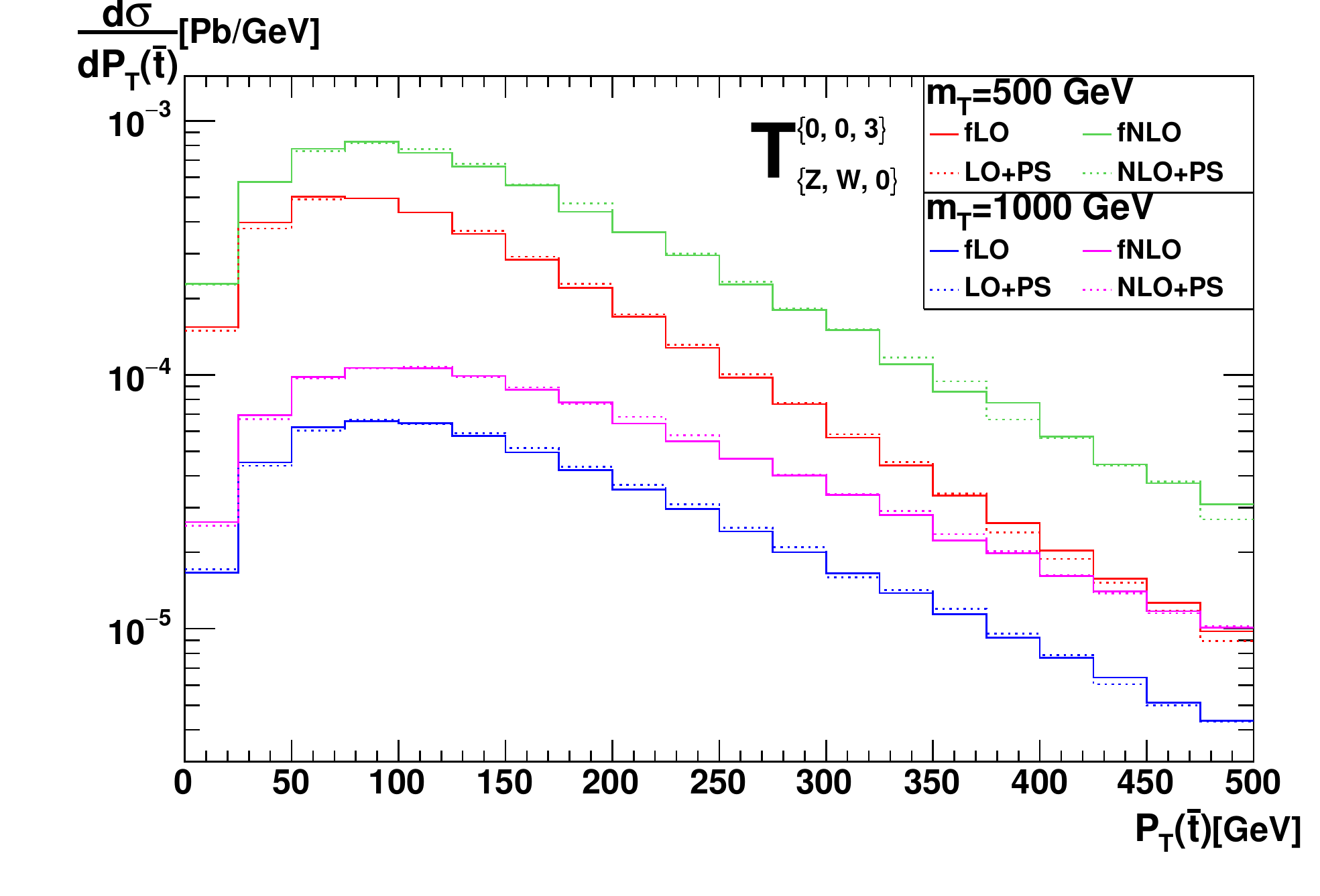}\includegraphics[width=7.5cm,height=5.5cm]{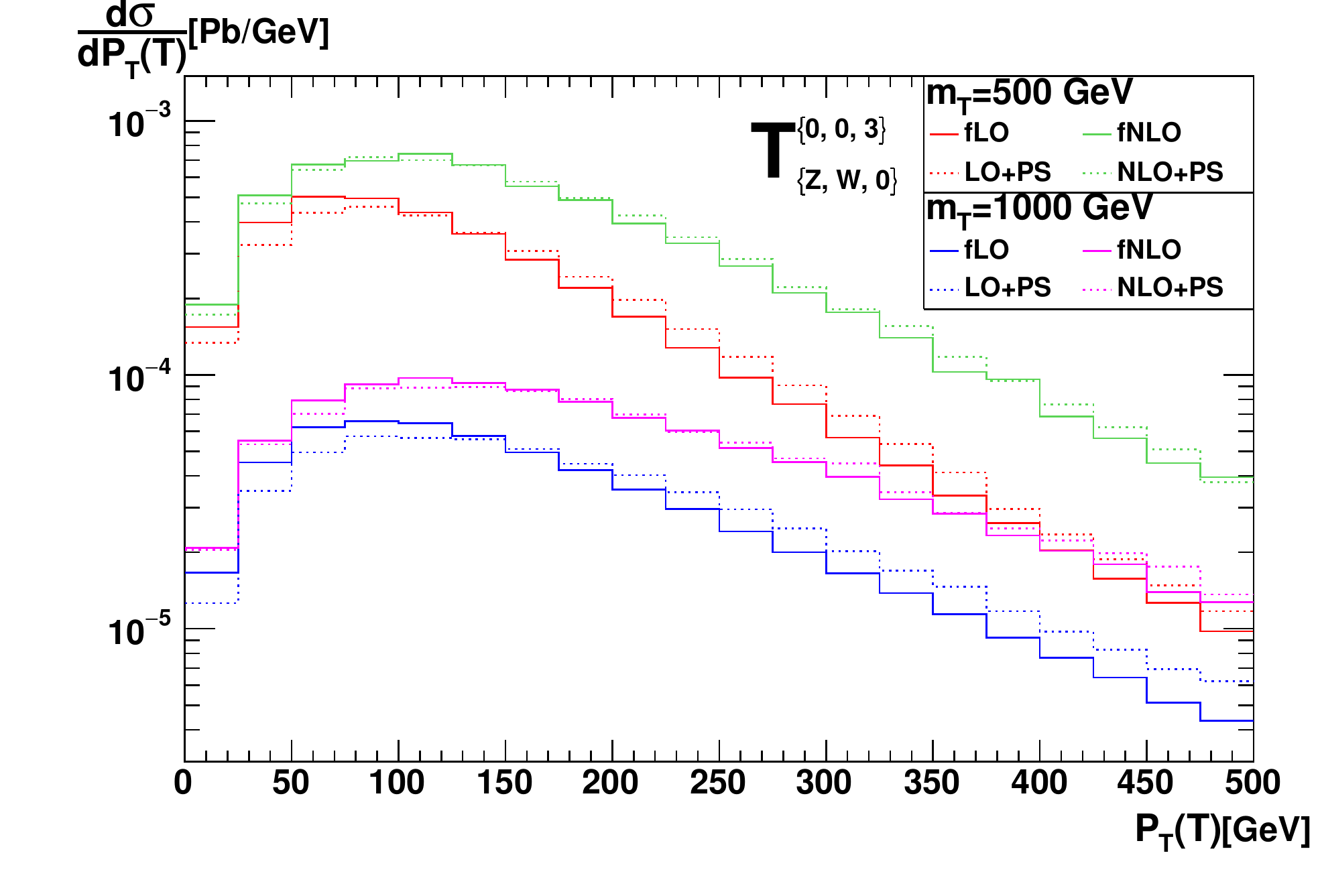}

\includegraphics[width=7.5cm,height=5.5cm]{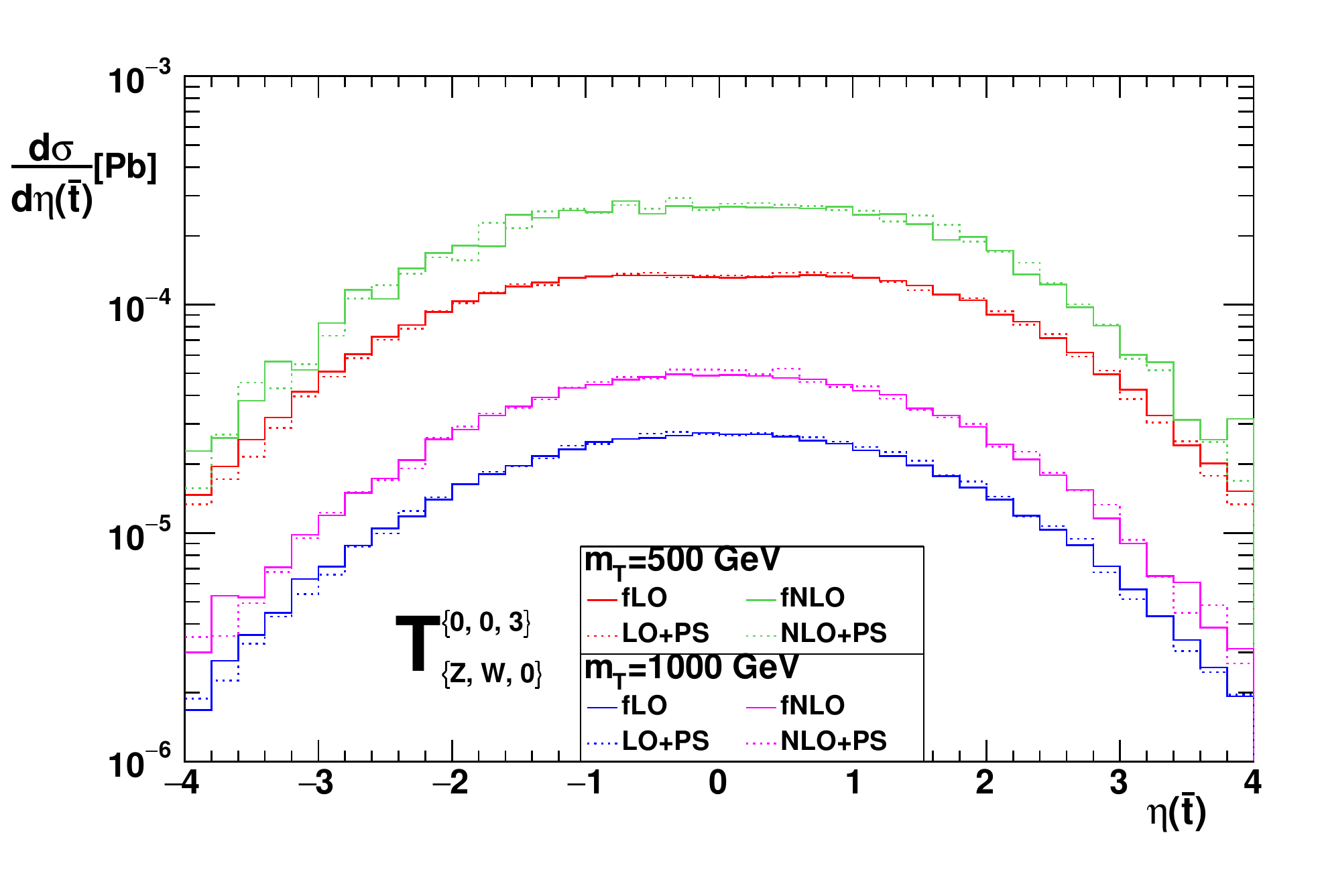}\includegraphics[width=7.5cm,height=5.5cm]{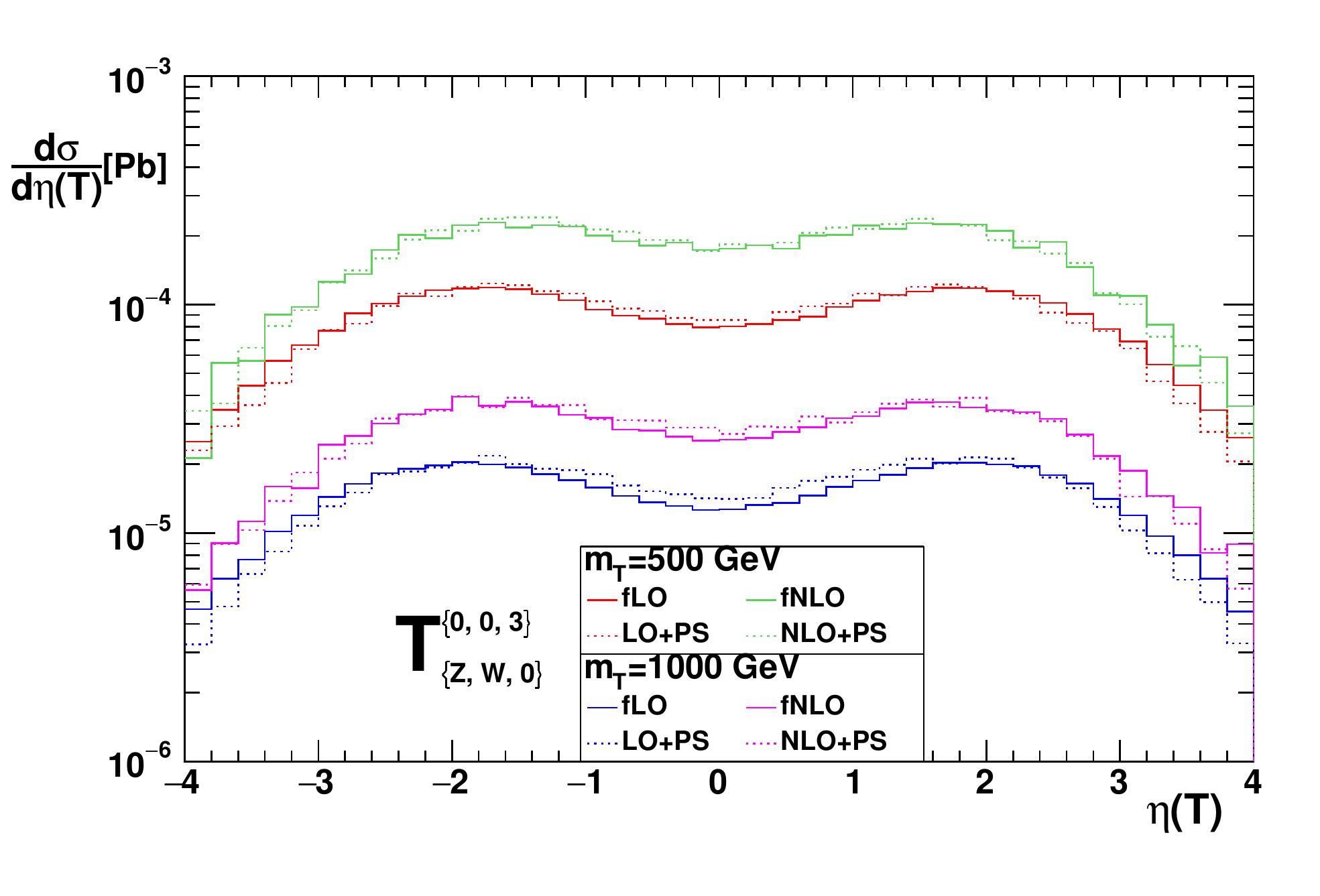}

\includegraphics[width=7.5cm,height=5.5cm]{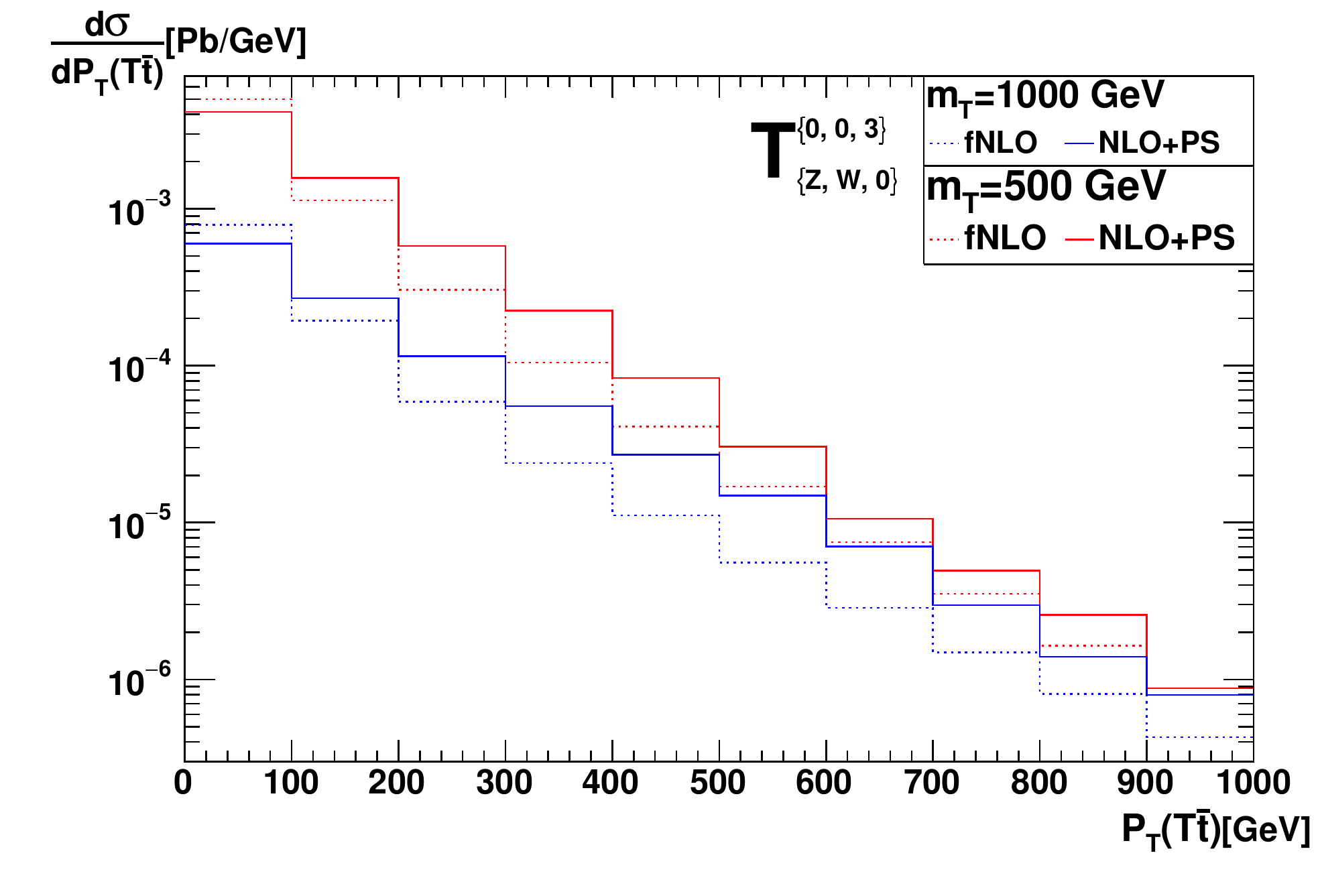}\includegraphics[width=7.5cm,height=5.5cm]{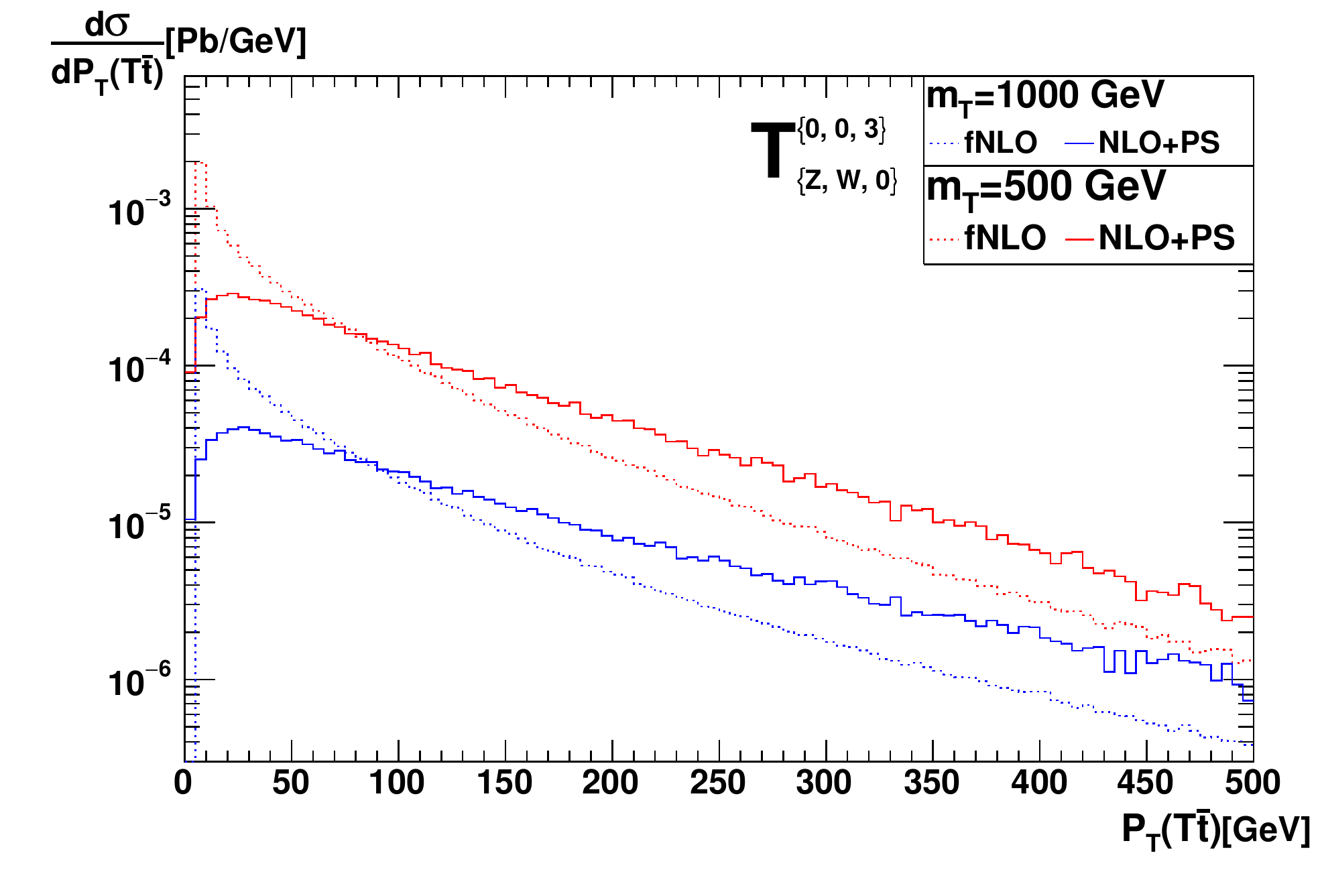}
}
\caption{\small Differential distributions for the benchmark scenario $\bf T^{\scriptscriptstyle\{0, 0, 3\}}_{\scriptscriptstyle\{Z, W, 0\}}$.}
\label{diffdisTt}
\end{figure}

%
In general, precise differential distributions are helpful to compare theory with experiment, and to put constraints on the free parameters of the model under consideration. In figure \ref{diffdisTt}, we show the differential distributions on the transverse-momentum and the pseudo-rapidity of $T$ and $\bar t$, and the transverse-momentum of the system $T\bar t$ for the benchmark scenario $\bf T^{\scriptscriptstyle\{0, 0, 3\}}_{\scriptscriptstyle\{Z, W, 0\}}$. These distributions are provided for fLO, fNLO, LO matched to parton shower (LO+PS) and NLO matched to parton shower (NLO+PS) for the two different values of the top partner mass $m_T=500, 1000\, \text{GeV}$, where {\tt Pythia8} \cite{ref41} is used. We employ {\tt MadAnalysis 5} \cite{madanalysis} to produce and analyse the distributions.  
We observe that the parton shower does not affect significantly the fLO and the fNLO transverse momentum and the rapidity distributions of the individual particles of the final state ($T$ and $\bar t$). Only for the very low $P_{\scriptscriptstyle T}$ and large $|\eta|$, considerable changes are seen which are nonetheless still bellow $28\%$. It tends to reduce the differential cross section at small $P_{\scriptscriptstyle T}$ and slightly increasing it at high $P_{\scriptscriptstyle T}$ for the single top partner, where the fNLO and the NLO+PS distributions match better compared to the LO and the LO+PS distributions in this region as expected (especially for low $T$ mass). The slight modifications of these distributions are attributed to the PDF uncertainties in this phase space region. On the other hand, the distributions of the anti-top quark show a good agreement between the fLO (fNLO) and the LO+PS (NLO+PS) for both the transverse momentum and the pseudo-rapidity, see the first four plots of figure \ref{diffdisTt}. By contrast, the spectrum of the NLO+PS transverse momentum distribution of the system $T\bar t$ ($P_{\scriptscriptstyle T}(T\bar t)$) shows a considerable change compared to fNLO predictions. 
In fact, the last two plots of figure \ref{diffdisTt} show the same $d \sigma / dP_{\scriptscriptstyle T}(T\bar t)$ spectrum. On the left side, it is given in bins of $100\, \text{GeV}$ up to $1000 \, \text{GeV}$ and, on the right side, it is provided in bins of $5\, \text{GeV}$ up to $500 \, \text{GeV}$. With the small binning, the fNLO predictions show a diverging behaviour for small $P_{\scriptscriptstyle T}(T\bar t)$. This is due to an incomplete compensation between the logarithmically divergent terms of the virtual and the real pieces in this region (due to the lack of phase space of the extra parton). However, with such a small binning the resummation to all orders is achieved in the NLO+PS approach, where the predictions are stable and well behaved for low transverse momentum. Since the fNLO and the NLO+PS are normalised to the same total cross section, one observes that at larger $P_{\scriptscriptstyle T}$ values NLO+PS predicts larger cross section since it gives lower predictions at low $P_T$. However it is expected that at very large $P_{\scriptscriptstyle T}$ they should converge as can be seen on the left plot (modulo the statistical fluctuations).
\section{Top partner production in association with top quark in 6FS}
\label{sec4}
\begin{figure}[tbp]
\centering
\includegraphics[width=14cm,height=3.5cm]{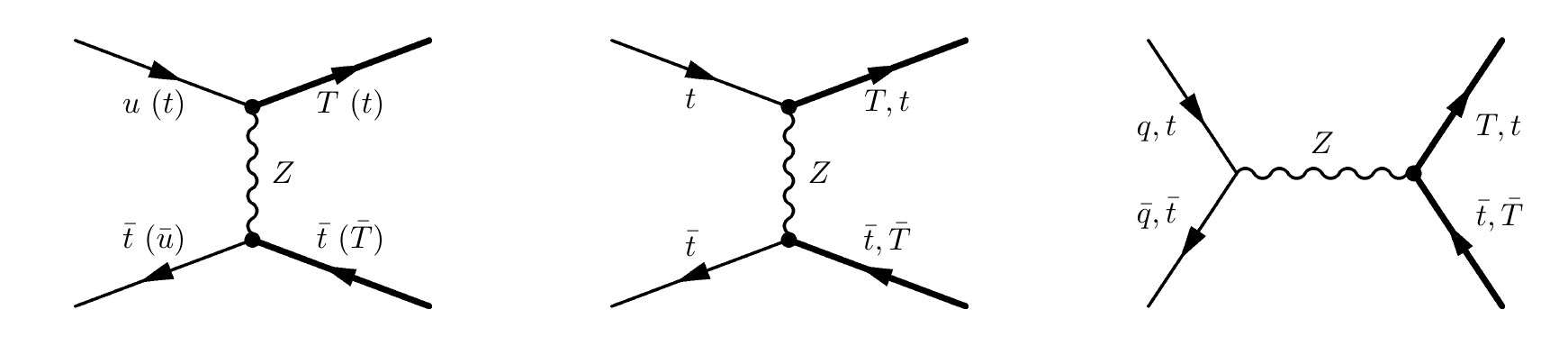}
    \caption{\footnotesize Leading-order Feynman diagrams for the reaction $pp\rightarrow T\bar t+\bar T t$ in 6FS.
    }
    \label{born6f}
   \end{figure}
\begin{figure}[tbp]
%
\centering
\subfloat[\label{sigma-mu-6fa}]{\includegraphics[width=6.5cm,height=5.0cm]{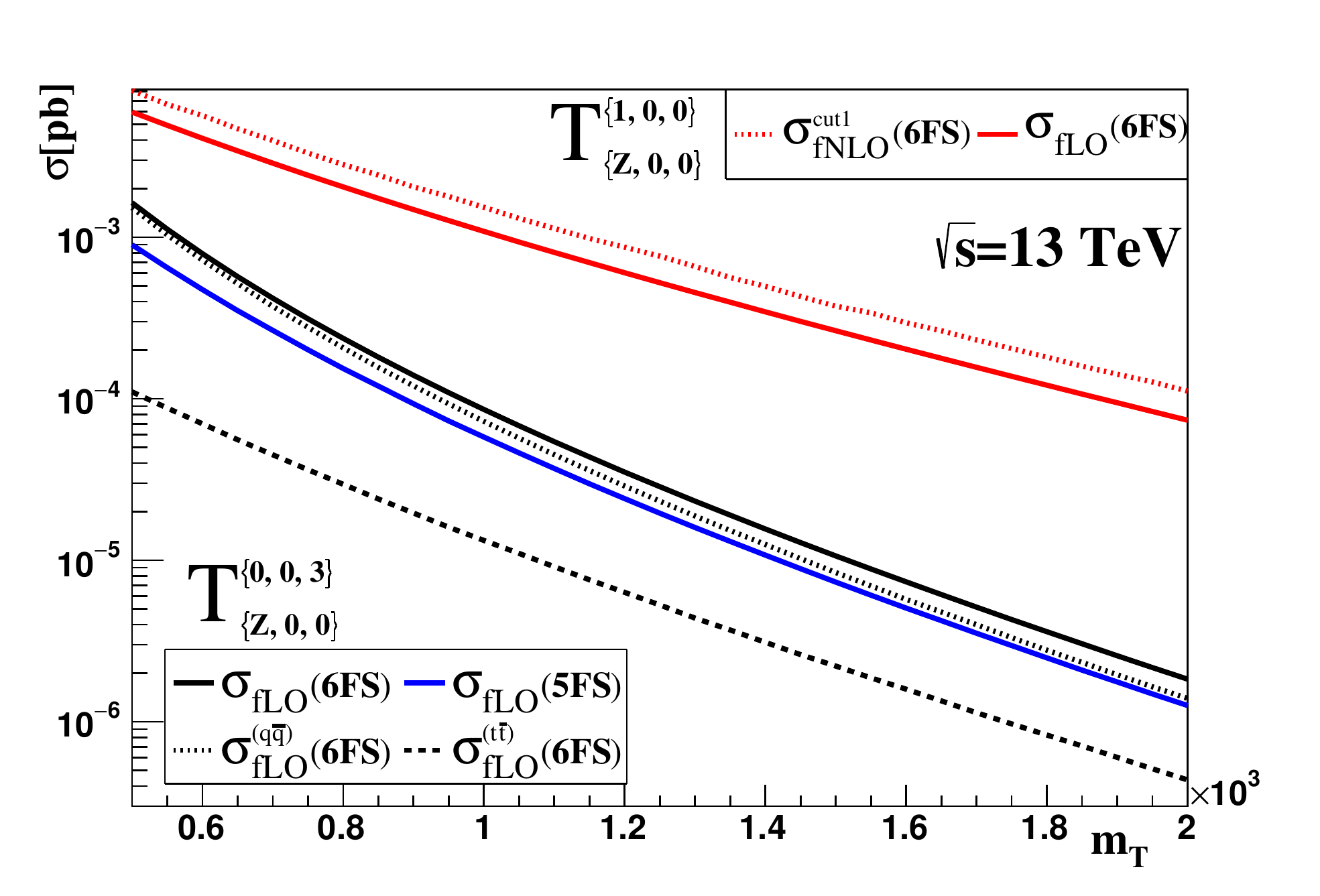}}\subfloat[\label{sigma-mu-6fb}]{\includegraphics[width=6.5cm,height=5.0cm]{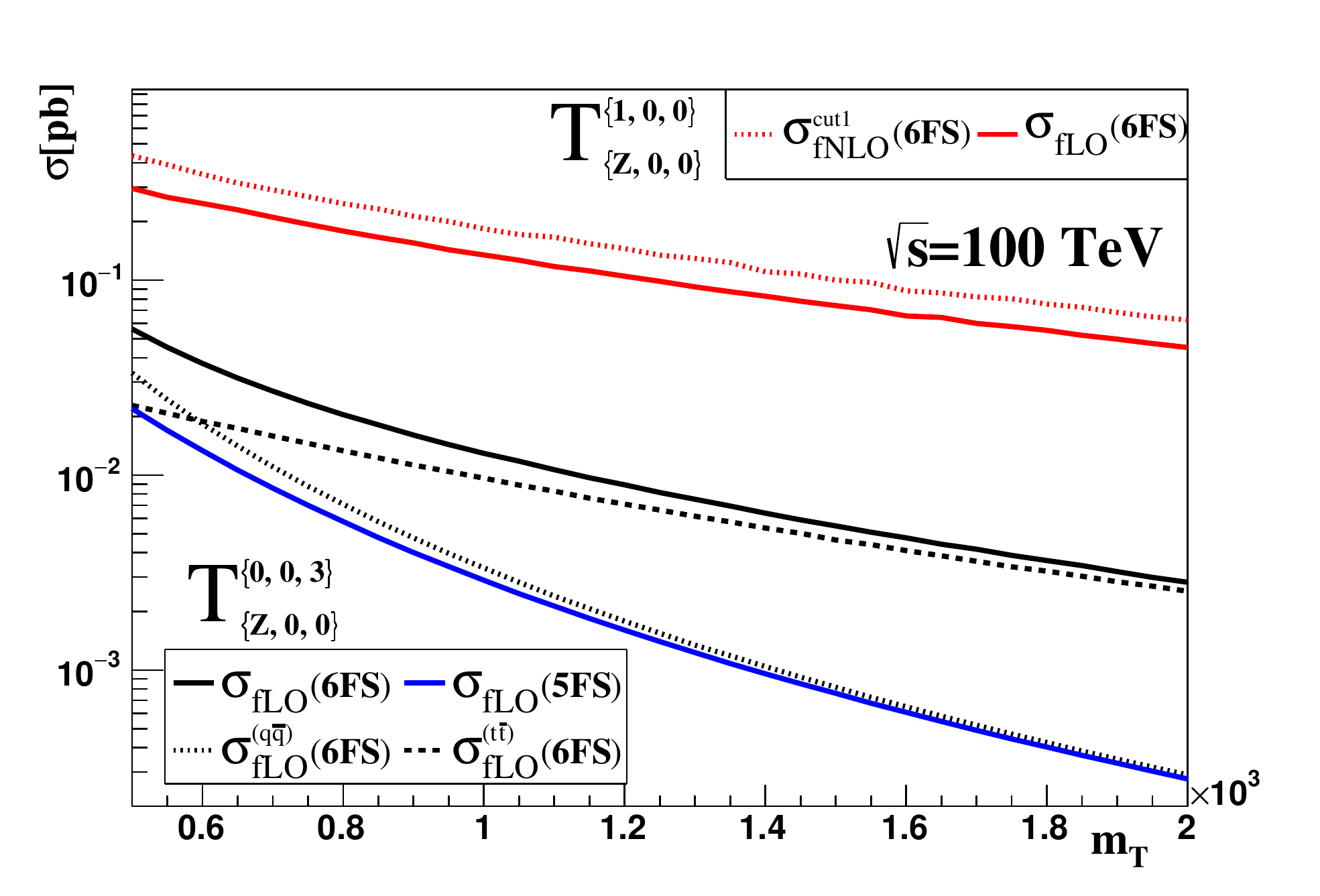}}
\caption{\small Variation of the cross section in term of $m_{\scriptscriptstyle T}$ at $\sqrt{s}=13\, \text{TeV}$ and $100\, \text{TeV}$}
\label{sigma-mu-6f}
\end{figure}

 
In the previous section, we have seen that benchmark scenarios without gauge boson $t$-channel exchange at the Born level have huge $K$-factors if no cut is applied on the extra partons of the real emission. To allow for such exchange (of $Z$ boson), one considers working in the 6FS. Furthermore, predictions in such scheme should be more reliable than in 5FS for very high energy regime $\sqrt{s}=100\, \text{TeV}$ (future proton-proton colliders). Also, 6FS leads to new production mechanisms of the same final state studied previously with more important cross sections than in 5FS.
In this section, we work in the standard massless 6FS, i.e. we take into account the top quark initiated processes where the top quark is treated as massless in the initial and the final states\footnote{To take into account the mass effects of the top quark explicitly, one has to modify the DGLAP evolution equations, extend
the subtraction method to include massive partons in the initial state $\cdots$ etc. Fortunately, neglecting the mass of the initial state heavy quarks, in our case, is justified. According to the simplified ACOT scheme (s-ACOT), the mass of the heavy quarks may be neglected for partonic sub-processes initiated by heavy quarks, see refs. \cite{6f1, 6f2, 6f3} for more detail.}. We suppose that the top quark partner can mix with quarks of the first or the third generations via the exchange of the $Z$ boson. In this case, the partonic processes which lead to the associated top partner with top quark final state are, 
\begin{align}
u\bar t\, (\bar ut)&\rightarrow T\bar t\, (\bar T t), & q\bar q&\rightarrow T\bar t+\bar T t, & t\bar t&\rightarrow T\bar t+\bar T t.
\label{process6f}
\end{align} 
where $q\equiv u, d, c, s, b, g$. The corresponding Feynman diagrams are depicted in figure \ref{born6f}. For the benchmark scenario $\bf T^{\scriptscriptstyle\{1, 0, 0\}}_{\scriptscriptstyle\{Z, 0, 0\}}$, only the first process contributes to the Born cross section. It is mediated by $t$-channel $Z$ boson (first diagram of figure \ref{born6f}). Regarding the benchmark scenario $\bf T^{\scriptscriptstyle\{0, 0, 3\}}_{\scriptscriptstyle\{Z, 0, 0\}}$, the second (same processes of 5FS) and the third processes contribute to the Born cross section. The latter one is initiated by top quarks and mediated by the exchange of $s$ and $t$-channels $Z$ boson (second and third diagrams of figure \ref{born6f}).
The calculation is done within the {\tt MadGraph} framework by making use of an UFO renormalised vector-like quark model, where the top quark is considered as massless. We use the fixed renormalization and factorization scales $\mu=\mu_F=\mu_R=m_T+m_t$, where $m_t$ is the physical top mass. We employ the PDF set {\tt NNPDF30\_nlo\_as\_0118\_nf\_6} for LO and NLO calculations. In table \ref{tab2}, we give the fLO and fNLO cross sections for $m_{\scriptscriptstyle T}=1000\, \text{GeV}$ and $m_{\scriptscriptstyle T}=1500\, \text{GeV}$, in the two benchmark scenarios, at $\sqrt{s}=13\,\text{TeV}$ and $\sqrt{s}=100\,\text{TeV}$. The fixed next-to-leading order cross section is provided with cut on the invariant mass of the top quark pair $M_{t\bar t}> 2m_t$ ($\sigma_{\scriptscriptstyle\text{fNLO}}^{\scriptscriptstyle\text{cut1}}$), and with cuts on the $\pt$ of the jet and $M_{t\bar t}> 2m_t$ ($\sigma_{\scriptscriptstyle\text{fNLO}}^{\scriptscriptstyle\text{cut2}}$)\footnote{The cuts on $\pt$ are chosen to reduce the $K$-factors. At $\sqrt{s}=13\, \text{TeV}$, we require that $p_T(j)>m_{\scriptscriptstyle T}/2$ for $m_{\scriptscriptstyle T}=1000\, \text{GeV}$ and $p_T(j)>2m_{\scriptscriptstyle T}/5$ for $m_{\scriptscriptstyle T}=1500\, \text{GeV}$. At $\sqrt{s}=100\, \text{TeV}$, we require that $p_T(j)>2m_{\scriptscriptstyle T}$ for the two mass values.}. In figure \ref{sigma-mu-6f}, we show the variation of the cross section in terms of the top partner mass.  
 \begin{table}[tbp]
\centering
 \renewcommand{\arraystretch}{1.40}
 \setlength{\tabcolsep}{12pt}
 \begin{adjustbox}{width=0.97\textwidth}
 \footnotesize
 \begin{tabular}{!{\vrule width 1pt}c|C{0.75cm}|C{0.75cm}!{\vrule width 1pt}l|l|l|l|C{0.5cm}!{\vrule width 1pt}}
  \noalign{\hrule height 1pt}
   \bf Scenario&\bf $m_{\scriptscriptstyle T}$ [GeV] &\bf$\sqrt{s}$ [TeV] &\bf $\sigma_{\scriptscriptstyle\text{fLO}}$ [pb]&\bf \bf $\sigma_{\scriptscriptstyle\text{fNLO}} \text{[pb]}$&\bf $\sigma_{\scriptscriptstyle\text{fNLO}}^{\scriptscriptstyle\text{cut1}} \text{[pb]}$&\bf $\sigma_{\scriptscriptstyle\text{fNLO}}^{\scriptscriptstyle\text{cut2}} \text{[pb]}$ &\bf $K${\footnotesize-factor}\\
  \noalign{\hrule height 1pt}
   \multirow{4}{*}{{\footnotesize\bf $\bf T^{\scriptscriptstyle\{0, 0, 3\}}_{\scriptscriptstyle\{Z, 0, 0\}}$}} & \multirow{2}{*}{\bf 1000} 
   &\bf 13
   &$(8.566\times 10^{-5}){}^{+2.4\%}_{-1.6\%}{}^{+2.0\%}_{-2.0\%}$ 
    &$(3.269\times 10^{-2}){}^{+20.5\%}_{-16.0\%}{}^{+2.4\%}_{-2.4\%}$
   &$(3.150\times 10^{-2}){}^{+20.6\%}_{-16.1\%}{}^{+1.8\%}_{-1.8\%}$ 
   &$(1.219\times 10^{-4}){}^{+23.8\%}_{-19.6\%}{}^{+3.6\%}_{-3.6\%}$&1.42\\
   && \bf 100
   &$(1.296\times 10^{-2}){}^{+45.7\%}_{-39.9\%}{}^{+1.2\%}_{-1.2\%}$
     &$(5.088\times 10^{0}){}^{+8.5\%}_{-7.7\%}{}^{+0.9\%}_{-0.9\%}$
   &$(4.997\times 10^{0}){}^{+10.1\%}_{-8.8\%}{}^{+0.9\%}_{-0.9\%}$
   &$(2.370\times 10^{-2}){}^{+18.1\%}_{-26.2\%}{}^{+1.8\%}_{-1.8\%}$&1.82\\
    \cdashline{2-8}
 & \multirow{2}{*}{\bf 1500} 
&\bf 13
&$(1.067\times 10^{-5}){}^{+1.0\%}_{-0.6\%}{}^{+3.3\%}_{-3.3\%}$
&$(6.195\times 10^{-3}){}^{+22.1\%}_{-17.0\%}{}^{+2.5\%}_{-2.5\%}$
&$(5.754\times 10^{-3}){}^{+22.8\%}_{-17.5\%}{}^{+3.0\%}_{-3.0\%}$
&$(1.694\times 10^{-5}){}^{+22.9\%}_{-19.2\%}{}^{+6.8\%}_{-6.8\%}$&1.59\\
&& \bf 100
&$(5.521\times 10^{-3}){}^{+39.8\%}_{-37.1\%}{}^{+1.6\%}_{-1.6\%}$
&$(2.396\times 10^{0}){}^{+9.9\%}_{-8.8\%}{}^{+0.9\%}_{-0.9\%}$
&$(2.131\times 10^{0}){}^{+10.5\%}_{-9.2\%}{}^{+0.9\%}_{-0.9\%}$ 
&$(9.669\times 10^{-3}){}^{+11.2\%}_{-18.5\%}{}^{+2.0\%}_{-2.0\%}$&1.75\\
\hline
\multirow{4}{*}{{\footnotesize\bf $\bf T^{\scriptscriptstyle\{1, 0, 0\}}_{\scriptscriptstyle\{Z, 0, 0\}}$}} 
   &\multirow{2}{*}{\bf 1000} &\bf 13
   &$(1.097\times 10^{-3}){}^{+18.2\%}_{-25.8\%}{}^{+1.4\%}_{-1.4\%}$ 
   &$(4.481\times 10^{-3}){}^{+13.3\%}_{-9.4\%}{}^{+1.2\%}_{-1.2\%}$
   &$(1.537\times 10^{-3}){}^{+28.0\%}_{-29.9\%}{}^{+1.4\%}_{-1.4\%}$&&1.40
   \\
   && \bf 100
   &$(1.364\times 10^{-1}){}^{+26.2\%}_{-30.9\%}{}^{+1.0\%}_{-1.0\%}$
   &$(3.811\times 10^{-1}){}^{+8.2\%}_{-6.9\%}{}^{+1.2\%}_{-1.2\%}$
   &$(1.931\times 10^{-1}){}^{+28.9\%}_{-29.2\%}{}^{+1.0\%}_{-1.0\%}$&&1.42
   \\
 \cdashline{2-8}
 &\multirow{2}{*}{\bf 1500} 
 &\bf 13
 &$(2.643\times 10^{-4}){}^{+11.9\%}_{-18.2\%}{}^{+2.6\%}_{-2.6\%}$
 &$(9.018\times 10^{-4}){}^{+18.0\%}_{-13.0\%}{}^{+3.0\%}_{-3.0\%}$
 &$(3.808\times 10^{-4}){}^{+22.4\%}_{-24.1\%}{}^{+2.3\%}_{-2.3\%}$&&1.44
 \\
 && \bf 100
 &$(7.423\times 10^{-2}){}^{+19.9\%}_{-24.1\%}{}^{+1.0\%}_{-1.0\%}$
 &$(1.820\times 10^{-1}){}^{+5.0\%}_{-5.3\%}{}^{+0.9\%}_{-0.9\%}$
 &$(1.037\times 10^{-1}){}^{+24.2\%}_{-24.5\%}{}^{+1.0\%}_{-1.0\%}$&&1.40 \\
\noalign{\hrule height 1pt}
\end{tabular}
\end{adjustbox}
\caption{\footnotesize fLO and fNLO cross sections in 6FS. For $\sigma_{\scriptscriptstyle\text{fNLO}}^{\scriptscriptstyle\text{cut1}}$, we apply a cut on the invariant mass $M_{t\bar t}>2m_t$. For $\sigma_{\scriptscriptstyle\text{fNLO}}^{\scriptscriptstyle\text{cut2}}$, we apply cuts on $p_T(j)$ and $M_{t\bar t}>2m_t$. The $K$-factor, given in the last column, is defined by the fractions: $\sigma_{\scriptscriptstyle\text{fNLO}}^{\scriptscriptstyle\text{cut1}}/\sigma_{\scriptscriptstyle\text{fLO}}$ for $\bf T^{\scriptscriptstyle\{1, 0, 0\}}_{\scriptscriptstyle\{Z, 0, 0\}}$ and $\sigma_{\scriptscriptstyle\text{fNLO}}^{\scriptscriptstyle\text{cut2}}/\sigma_{\scriptscriptstyle\text{fLO}}$ for $\bf T^{\scriptscriptstyle\{0, 0, 3\}}_{\scriptscriptstyle\{Z, 0, 0\}}$.}
\label{tab2}
\end{table}

 We observe that, $\sigma_{\scriptscriptstyle\text{fLO}}$ in the scenario $\bf T^{\scriptscriptstyle\{1, 0, 0\}}_{\scriptscriptstyle\{Z, 0, 0\}}$ (the reaction is allowed only in 6FS) is larger than $\sigma_{\scriptscriptstyle\text{fLO}}$ in the scenario $\bf T^{\scriptscriptstyle\{0, 0, 3\}}_{\scriptscriptstyle\{Z, 0, 0\}}$ in both 5FS and 6FS at $\sqrt{s}=13\, \text{TeV}$ (solid red, black and blue curves in sub-figure \ref{sigma-mu-6fa}). This shows the dominance of the $t$-channel exchange even if the  luminosity of the top is very small compared to the other quarks at the LHC. In fact, the virtuality of the $t$-channel exchanged $Z$ boson can be very small, which explains the increase of the cross section. The same thing is observed for very high center-of-mass energy, see sub-figure \ref{sigma-mu-6fb}. 
 
  We find that, the fLO cross section of $\bf T^{\scriptscriptstyle\{0, 0, 3\}}_{\scriptscriptstyle\{Z, 0, 0\}}$ in the 6FS is very close to the cross section calculated in the 5FS at $\sqrt{s}=13\,\text{TeV}$, see the solid black and blue curves in sub-figure \ref{sigma-mu-6fa}. This is due to the fact that the leading contribution (second process in eq. (\ref{process6f})) is the same as in the 5FS, and the sub-leading contribution (third process in eq. (\ref{process6f})) is very small compared to the former one because of the very small top quark PDF at the LHC. In fact, the slight difference between the cross sections in the two schemes is due mainly to the top mass which is neglected in 6FS. We observe also that the initiated top quark sub-processes contribution increases with the mass of the top partner, see the black dotted and dashed curves in sub-figure \ref{sigma-mu-6fa}. It represents, respectively, about $7\%$ and  $24\%$ of the total cross section for $m_{\scriptscriptstyle T}=500\, \text{GeV}$ and $m_{\scriptscriptstyle T}=2000\, \text{GeV}$ at $\sqrt{s}=13\, \text{TeV}$. For very high center-of-mass energy (sub-figure \ref{sigma-mu-6fb}), it becomes the dominant contribution especially for $m_{\scriptscriptstyle T}$ above $600\, \text{GeV}$. It represents about $41\%$ and $90\%$ of the total cross section for $m_{\scriptscriptstyle T}=500\, \text{GeV}$ and $m_{\scriptscriptstyle T}=2000\, \text{GeV}$, respectively.
  This shows the increase of the top density inside the proton with the increase of the  mass scale and for very high centre-of-mass energy. It shows also, the dominance of the $t$-channel exchange $Z$ boson contribution (second diagram of figure \ref{born6f}).
 
 In what concerns the $K$-factors. They are still very large for the benchmark scenario $\bf T^{\scriptscriptstyle\{0, 0, 3\}}_{\scriptscriptstyle\{Z, 0, 0\}}$ if no cut is applied. However, if we apply cuts on the invariant mass of the system $t\bar{t}$ and $\pt$ of the jet, they are significantly reduced, see the first part of the last column of table \ref{tab2}. Regarding the scenario $\bf T^{\scriptscriptstyle\{1, 0, 0\}}_{\scriptscriptstyle\{Z, 0, 0\}}$, the $K$-factors (without cut) are not very large like in the previous case (about $3.41$ at $\sqrt{s}=13\, \text{TeV}$ and $2.45$ at $\sqrt{s}=100\, \text{TeV}$ for $m_{\scriptscriptstyle T}=1500\, \text{GeV}$). If we restrict $M_{t\bar t}>2m_{t}$, they are considerably reduced as shown in the last column of table \ref{tab2}. Actually, these cuts enforce the virtuality of the $t$-channel exchanged $Z$ boson for the gluon-quark (anti-quark) real emission to be larger which reduces their contribution. 
\section{Investigation of the single $T$ decaying into a Higgs and a top quark}
\label{sec5}
Following the benchmark scenarios discussed above, the top quark partner can decay into one of the third generation SM quarks via one of the three channels $bW$, $tZ$ or $tH$, where mixing with the first and the second quark generations is ignored. In this section, we examine the $t\bar t H$ final state obtained from the reaction studied above (in 5FS) through  the decay of the heavy vector-like quark into a top quark and a Higgs boson ($pp\longrightarrow T\bar t+\bar T t\longrightarrow t\bar t H$) at $\sqrt{s}=13\, \text{TeV}$. We do not consider the full NLO calculation, but we just match the calculation done in section \ref{sec3} to parton-shower ({\tt pythia8}) and perform the decay by {\tt MadSpin} \cite{ref42} and {\tt MadWidth} \cite{madwidth}. This approximation constitutes a first step towards a more sophisticated treatment, where we are going to consider the full next-to-leading order calculation for $t\bar t H$  production at the LHC for models with vector-like quarks.   

The investigation of this final state, in SM and BSM, has received more and more attention by both theorists and experimentalists in the last few years, especially after the discovery of the Higgs boson. In the context of confirming the predicted and the measured properties of the SM Higgs boson, and aiming to discover BSM physics, ATLAS and CMS collaborations analyse, in many recent publications, the $t\bar{t}H$ final state for different decay modes at the LHC {\tt runI} and {\tt runII}. For example, in \cite{ref31} evidence for the production of the Higgs boson in association with a top quark pair at ATLAS were shown. They presented a good agreement with the measured and the predicted cross sections for different Higgs decay modes and data collected  from LHC {\tt runI} and {\tt runII}. Alternative publications by both collaborations, on the same subject, are given in \cite{ref32, ref33, ref35, ref37}.
The search for the single production of top quark partner $T$ decaying into a Higss boson and top quark was the subject of many publications of ATLAS and CMS collaborations. In \cite{THt2} for example, CMS puts for the first time exclusion limits on the cross section for $T$ single production at $\sqrt{s}=13\, \text{TeV}$, where they targeted the decay of the Higgs boson into a pair of bottom quarks, and the decay of the top quark includes a muon or an electron. The other possible decay modes ($tZ$ and $bW$) of the singly produced heavy top quark partner were investigated in \cite{TZt1, TZt2, TWb1, TWb2} for example.

At this stage, we focus on the differential $K$-factors, i.e. the ratio of the NLO+PS and LO+PS differential cross sections. Let's consider the differential distributions of the following observables: the transverse momentum of the Higgs boson ($\pt(H)$), the transverse momentum of the top quark ($\pt(t)$), the invariant mass of the final state system $t\bar tH$ ($M_{t\bar t H}$), the rapidity of the Higgs ($Y_{\scriptscriptstyle H}$), and the rapidity of the top quark ($Y_{\scriptscriptstyle t}$). The distributions are calculated at LO and NLO matched to parton shower for the benchmark scenario ${\bf T^{\scriptscriptstyle\{0, 0, 3\}}_{\scriptscriptstyle\{Z, W, H\}}}$. The calculation is done at center-of-mass energy $\sqrt{s}=13\, \text{TeV}$  for the two different mass values $m_{\scriptscriptstyle T}=500\, \text{GeV}$ and $m_{\scriptscriptstyle T}=800\, \text{GeV}$, where the dynamical scale is used (the average transverse mass of the final state particles). 

\begin{figure}[tbp]
\centering
\subfloat[\label{a}]{\includegraphics[width=7.75cm,height=5.5cm]{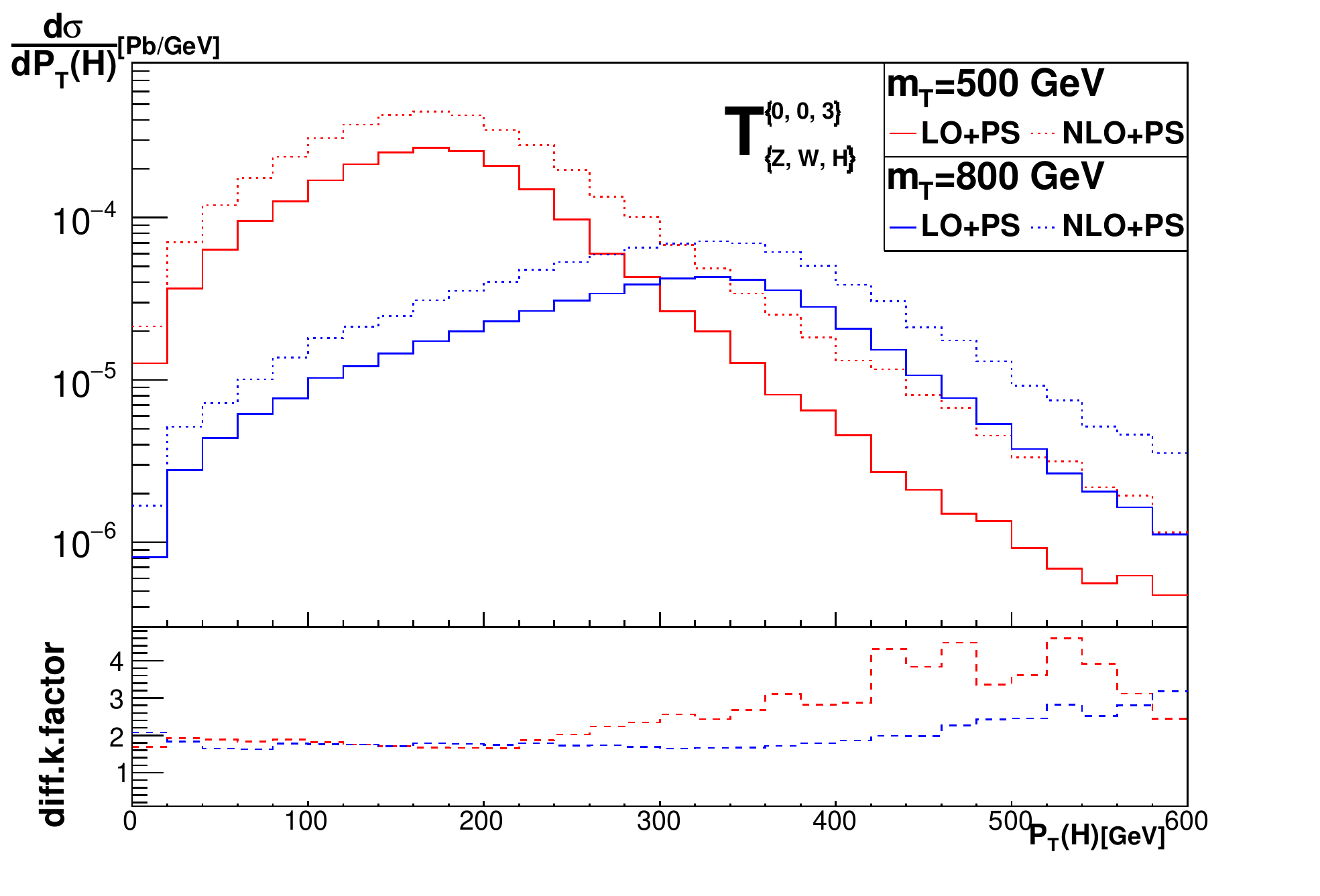}}
\subfloat[\label{b}]{\includegraphics[width=7.75cm,height=5.5cm]{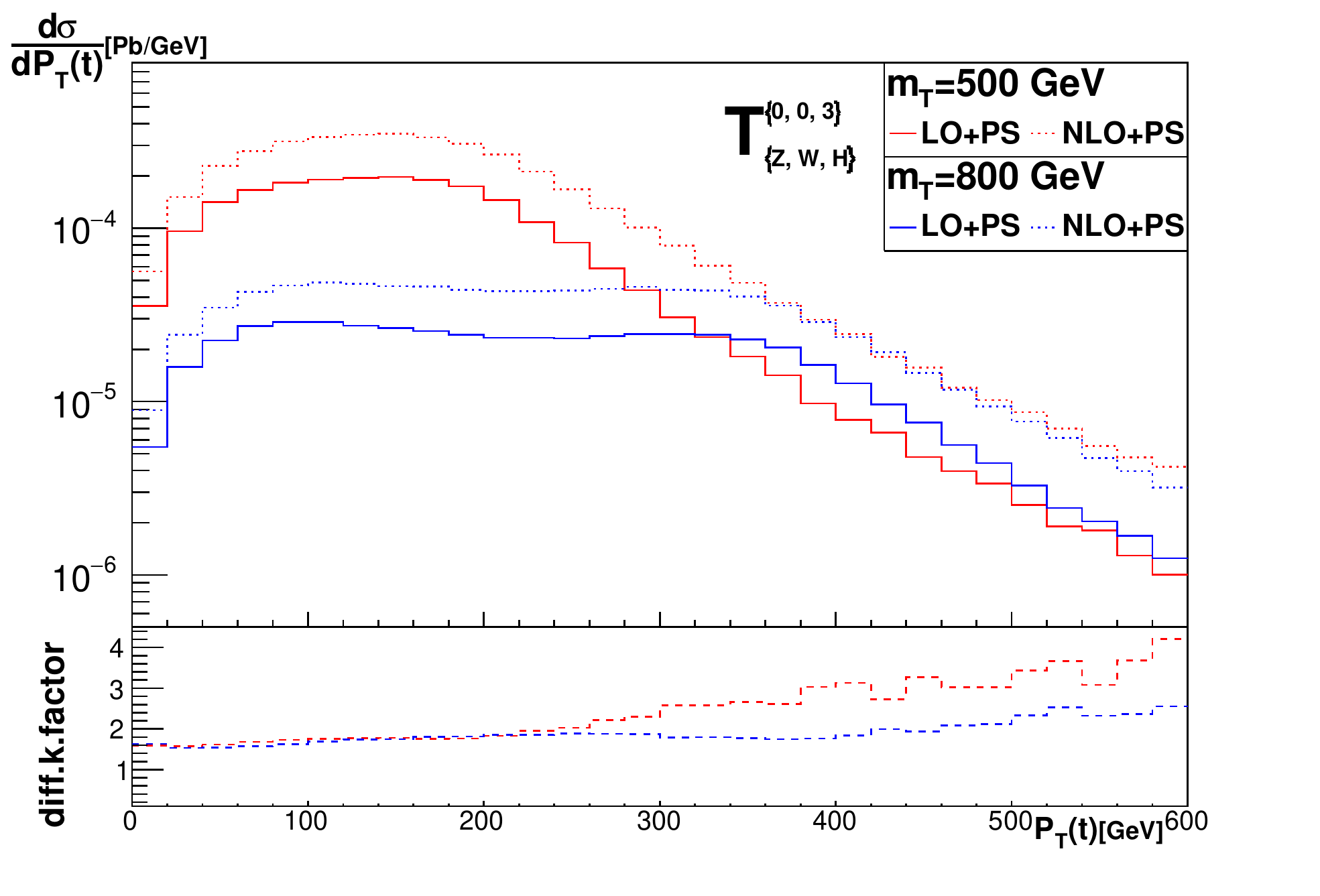}}

\vspace{-0.4cm}
\subfloat[\label{c}]{\includegraphics[width=7.75cm,height=5.5cm]{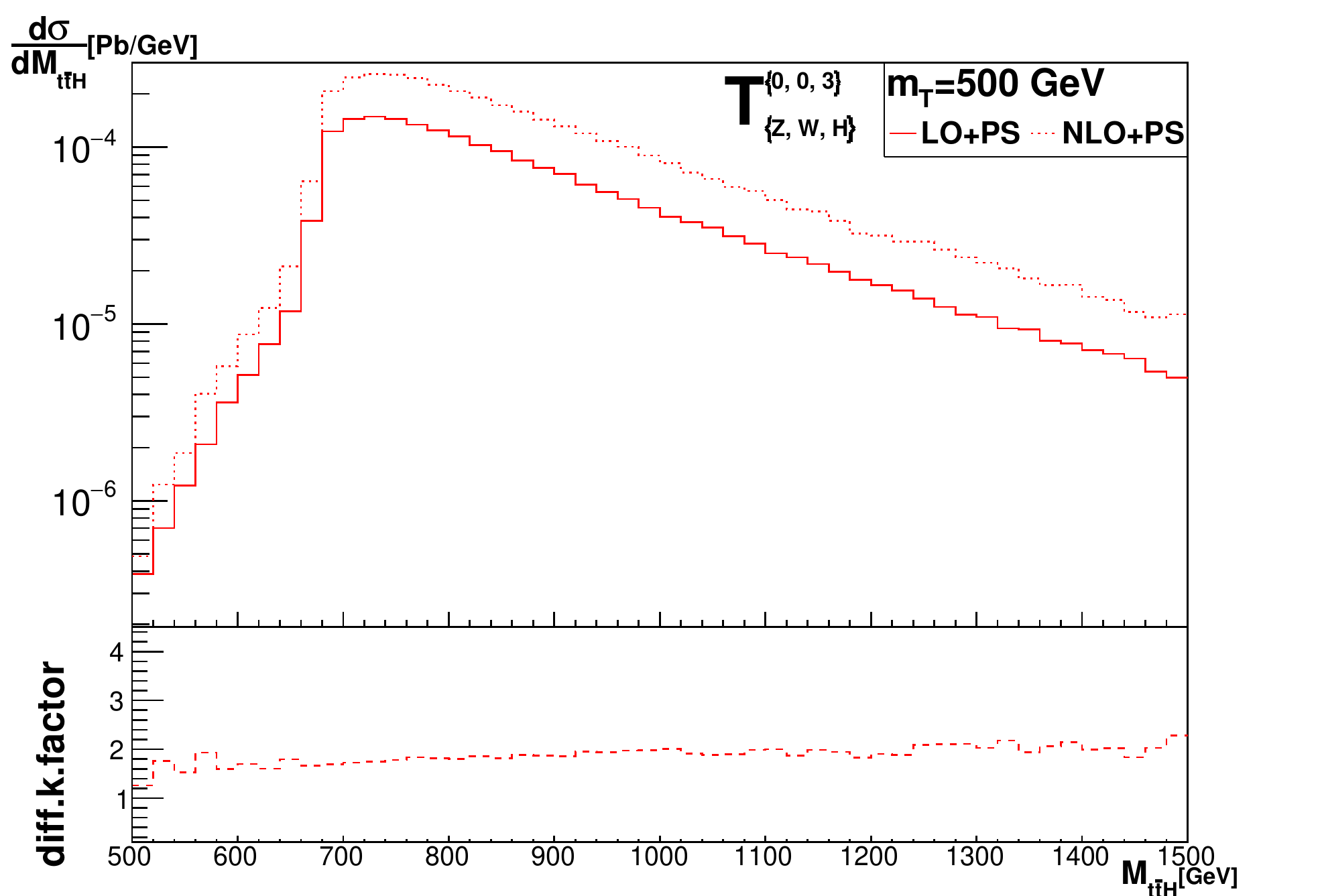}} 
\subfloat[\label{d}]{\includegraphics[width=7.75cm,height=5.5cm]{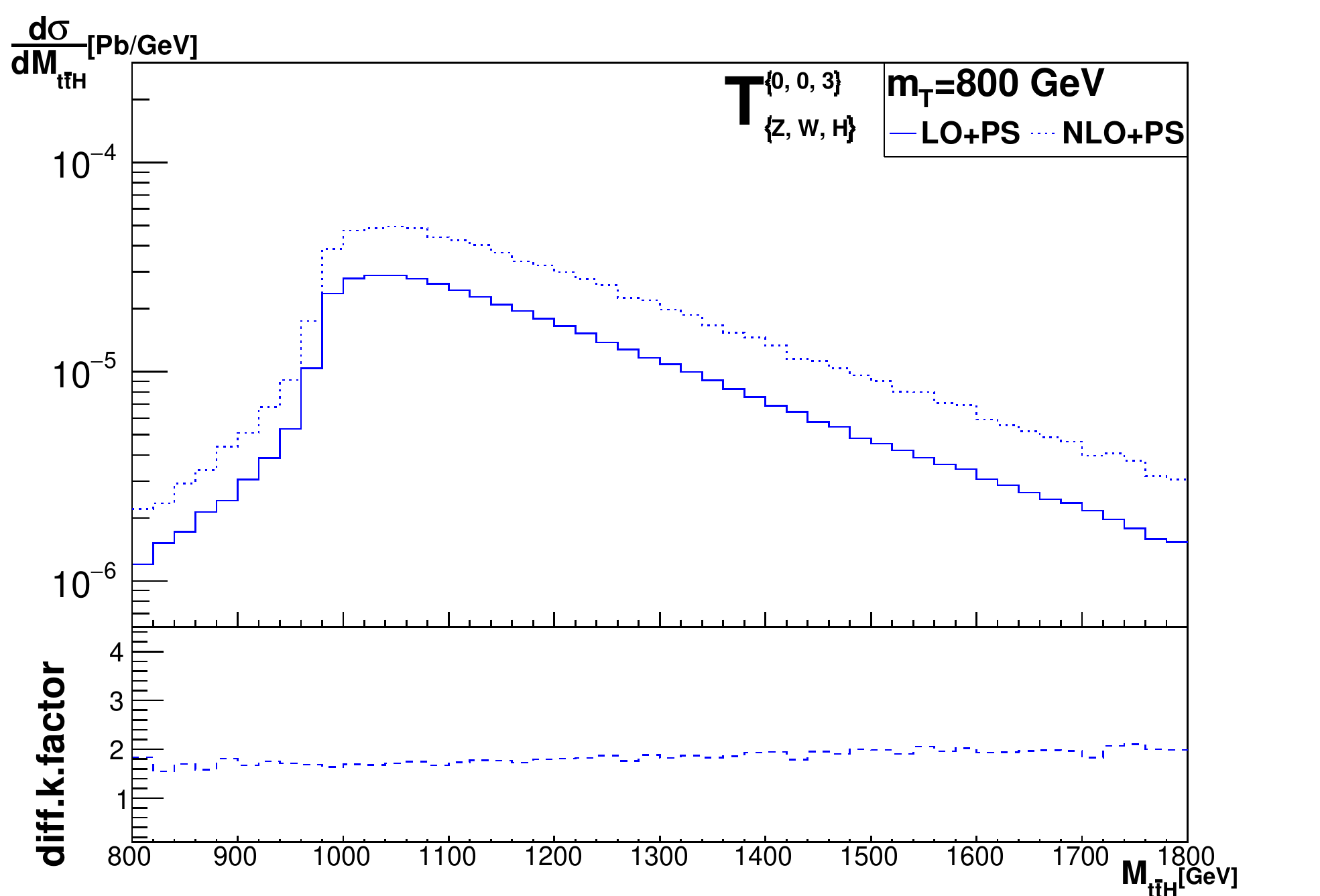}} 

\vspace{-0.4cm}
\subfloat[\label{e}]{\includegraphics[width=7.75cm,height=5.5cm]{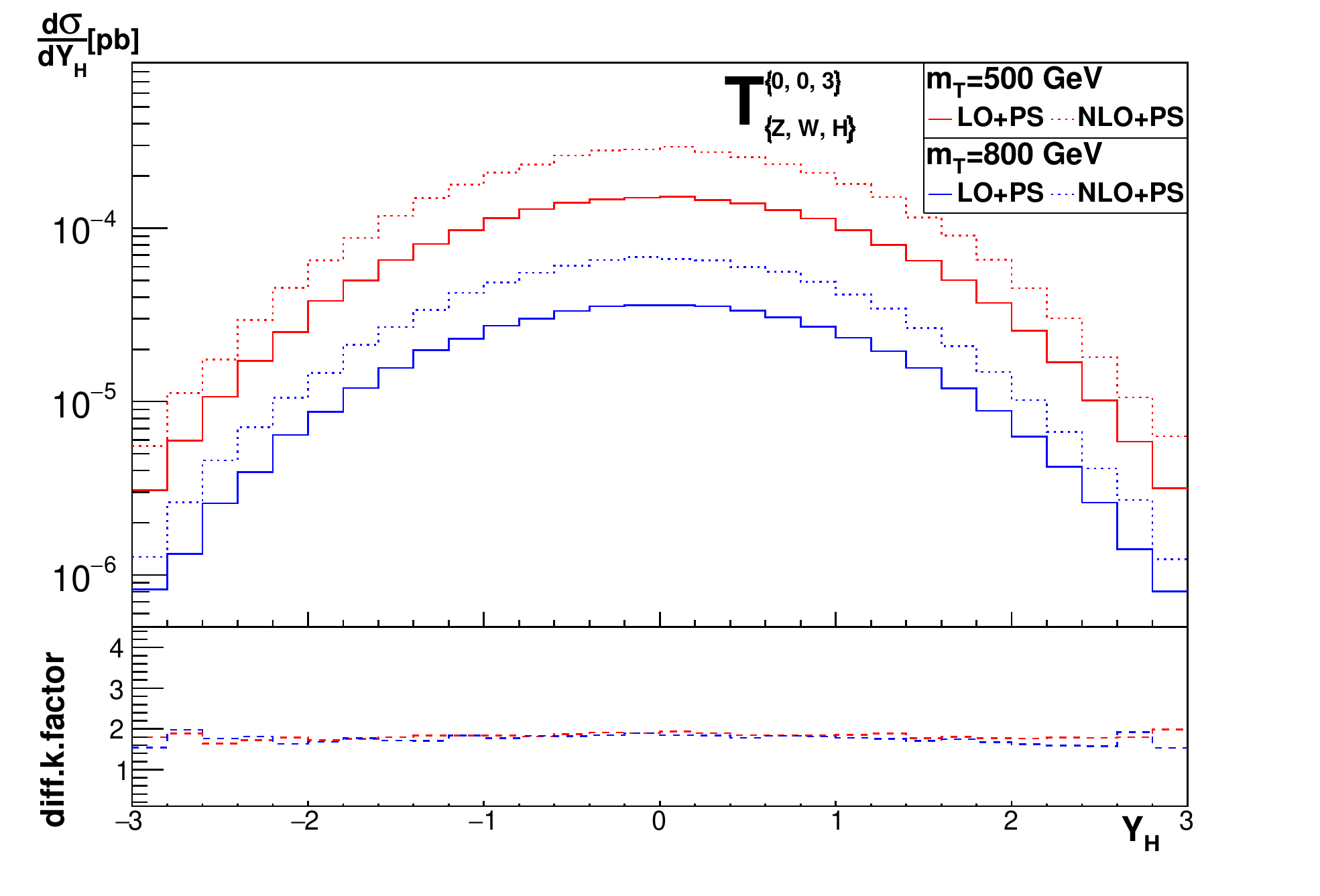}} 
\subfloat[\label{f}]{\includegraphics[width=7.75cm,height=5.5cm]{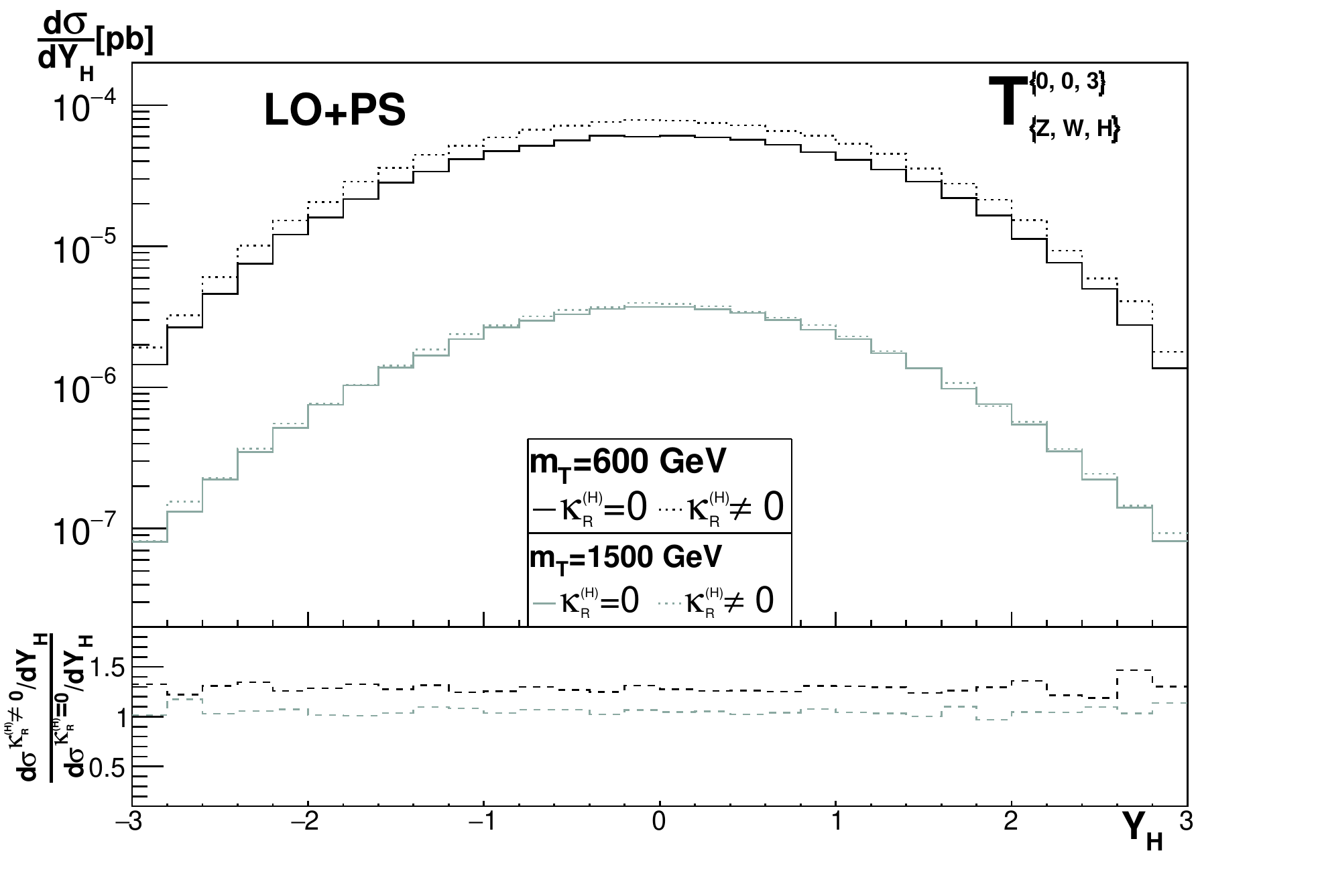}} 
  \caption{\footnotesize Differential distributions of some observables for the $t\bar t H$ final state in the benchmark scenario ${\bf T^{\scriptscriptstyle\{0, 0, 3\}}_{\scriptscriptstyle\{Z, W, H\}}}$, for different $m_{\scriptscriptstyle T}$ values, at leading and next-to-leading orders matched to parton shower.}
\label{difttbarh}
\end{figure}

In figures \ref{a}, \ref{b}, \ref{c}, \ref{d} and \ref{e}, we plot the differential cross sections as functions of the different kinematic variables, introduced above. The solid red and blue plots represent the LO+PS distributions and the dotted red and blue plots represent the NLO+PS distributions. In the lower panels, we display the differential $K$-factors for each distribution.  
The first comment is that all the LO+PS and NLO+PS distributions, for the same mass, have mainly identical shapes with different magnitudes, where the magnitudes of the shown observables decrease if the top partner mass increase, as expected\footnote{For large $P_{\scriptscriptstyle T}(t)$ and $P_{\scriptscriptstyle T}(H)$, it is not the case. However, in this region, the differential cross sections are very small and dominated by the statistical fluctuations.}. Actually, we have found that the integrated cross sections for $m_{\scriptscriptstyle T}=500\, \text{GeV}$ are: $\sigma^{\text{LO+PS}}=2.15\times 10^{-3}\,\text{pb}$ and $\sigma^{\text{NLO+PS}}=3.97\times 10^{-3}\,\text{pb}$, and for $m_{\scriptscriptstyle T}=800\, \text{GeV}$ are: $\sigma^{\text{LO+PS}}=5.15\times 10^{-4}\,\text{pb}$ and  $\sigma^{\text{NLO+PS}}=9.25\times 10^{-4}\,\text{pb}$. This implies that the $K$-factors are of order $1.80$, i.e. they are not changed and they are still large for this final state. This is due to the contribution of the gluon-quark (anti-quark) real emission Feynman diagrams as explained in the previous section. We observe that the differential $K$-factors of the Higgs and the top quark transverse momentum are relatively flat for $P_T$ ranging between  $0$-$250\, \text{GeV}$ and between $0$-$420\, \text{GeV}$ for $m_{\scriptscriptstyle T}=500\, \text{GeV}$ and $m_{\scriptscriptstyle T}=800\, \text{GeV}$, respectively. They coincide for $P_T$ ranging between $0$-$250\, \text{GeV}$ for both distributions and for the two masses, where they are about $1.9$. For higher $P_T$'s ($P_T>250$ for the smaller mass and $P_T>420$ for the larger mass), the $K$-factors become arbitrary increasing, i.e. the higher order corrections are much more important than the leading order in this region. The invariant mass distributions of the final state system peak at $750\, \text{GeV}$ for $m_T=500$, and at $1150\, \text{GeV}$ for $m_T=800\, \text{GeV}$ and decrease very quickly with increasing $M_{t\bar t h}$. The differential $K$-factors are relatively flat (around 2.0) with some discrepancy before the peaks and for very high invariant mass. Regarding, the differential $K$-factor for the Higgs rapidity spectrum, it is flat and sable (nearly 2.0) except for high $|y_{\scriptscriptstyle H}|$, where it shows a little perturbation. Anyway, in the regions where the differential $K$-factors are not stable, the differential distributions are dominated by the PDF uncertainties (and statistical fluctuations).
 
 Finally, we remind that for the decay of the vector-like top partner to a Higgs and a top quark, we have taken into account the sub-leading right-handed coupling $T$-$H$-$t$. The later one has non-negligible effects if the mass of $T$ is lower than $1000\, \text{GeV}$. In figure \ref{f}, we display the LO+PS differential cross section as a function of the Higgs rapidity for $m_{\scriptscriptstyle T}=600\, \text{GeV}$ and $m_{\scriptscriptstyle T}=1500\, \text{GeV}$, and for both cases $\khr=0$ and $\khr\neq0$. In the lower panel, we plot the ratio of this distribution for $\khr=0$ and $\khr\neq0$, i.e. ($d\sigma({\footnotesize{\khr\neq0}})/dY_{\scriptscriptstyle H})/(d\sigma({\footnotesize\khr=0})/dY_{\scriptscriptstyle H}$). This differential fraction is about $1.28$ for $m_{\scriptscriptstyle T}=600\, \text{GeV}$ and $1.05$ for $m_{\scriptscriptstyle T}=1500\, \text{GeV}$ (for most of the bins), which is in agreement with the integrated cross section. For the later one, we have found that $\sigma(\khr\neq0)=1.14\times10^{-3}\,\text{pb}$ and $\sigma(\khr=0)=8.94\times10^{-4}\,\text{pb}$ for $m_{\scriptscriptstyle T}=600\, \text{GeV}$, and $\sigma(\khr\neq0)=5.18\times10^{-5}\,\text{pb}$ and $\sigma(\khr=0)=4.94\times10^{-5}\,\text{pb}$ for $m_{\scriptscriptstyle T}=1500\, \text{GeV}$. Then, the cross section is about $28\%$ ($5\%$) higher if $\khr$ is included for $m_{\scriptscriptstyle T}=600\, \text{GeV}$ ($m_{\scriptscriptstyle T}=1500\, \text{GeV}$).  
\section{Conclusion}
In this paper, we have presented the calculation of the QCD next-to-leading order cross section for the production of a single vector-like top partner in association with a top quark, for many possible benchmark scenarios, in proton-proton collisions, in the five-flavour and the six-flavour schemes. The computation is performed within the {\tt MadGraph5\_aMC@NLO} framework by making use of 5FS and 6FS UFO NLO vector-like quarks models, which offers all the required ingredients to do such calculation. We have investigated the fixed-order and the fixed-order matched to parton shower differential distributions at LO and NLO accuracies. We have examined the decay of the heavy top partner into a Higgs boson and a top quark which leads to the interesting Higgs with a top quark pair final state. This constitutes a first step towards performing the full next-to-leading order calculation of the production of the associated top quark pair with the Higgs at the LHC, including all the QCD and the mixed QCD and electro-weak effects, for models with vector-like quarks. 

We have shown, in 5FS, that the NLO $K$-factors for the total cross section and the differential distributions are very different from one benchmark scenario to another. For the benchmark scenarios where the heavy top quark partner interacts with the third generation quarks and the $Z$ boson (and the Higgs), the $K$-factors are gigantic and increase dramatically with the mass of the top partner. We have found that this is due mainly to the $t$-channel gluon-quark (anti-quark) real emission which appears exclusively at NLO order, and it is seen as a new channel which dominates over the $s$-channel tree level and one loop Feynman diagram contributions. We have shown that if we enforce the transverse momentum of the extra partons of the real emission to be larger than a given fraction of $m_{\scriptscriptstyle T}$, the $K$-factors are significantly reduced. Regarding the scenarios where the heavy top quark partner interacts with the third generation quarks and the $W$ boson (and the Higgs), the $K$-factor is about one however the bottom quark PDF uncertainties are relatively large. On the other hand, the scenarios where the heavy top quark partner interacts with the third generation quarks, the $Z$ and $W$ bosons (and the Higgs), the $K$-factor is about two, which is due to the gluon-quark (anti-quark) contribution.

We have found that the Born cross section, in 6FS, if $T$ is assumed to mix only with first quark generation (the reaction is allowed only in 6FS) is larger than the one obtained in the scenario ${\bf T^{\scriptscriptstyle\{0, 0, 3\}}_{\scriptscriptstyle\{Z, 0, 0\}}}$ for the same final state. We have shown  that top quark initiated processes contribution becomes more and more important for high mass scale and for very high center-of-mass energy (in ${\bf T^{\scriptscriptstyle\{0, 0, 3\}}_{\scriptscriptstyle\{Z, 0, 0\}}}$). It becomes dominant for $T$ mass above $600\, \text{GeV}$ at $\sqrt{s}=100\, \text{TeV}$. This is explained by the increase of the top quark density inside the proton in this regime and, the dominance of the $t$-channel $Z$ boson exchange. In what concern the $K$-factors in 6FS, they are still large if no cut is applied for the case ${\bf T^{\scriptscriptstyle\{0, 0, 3\}}_{\scriptscriptstyle\{Z, 0, 0\}}}$ like in 5FS. However, for ${\bf T^{\scriptscriptstyle\{1, 0, 0\}}_{\scriptscriptstyle\{Z, 0, 0\}}}$ they are significantly lower than in ${\bf T^{\scriptscriptstyle\{0, 0, 3\}}_{\scriptscriptstyle\{Z, 0, 0\}}}$ even if without phase space restriction especially for very high energy regime. 

We mention that we have included the right-handed chirality of the coupling of the Higgs boson to the top partner and the top quark. We have shown, in the last section, that it is very important to include such coupling for low top partner mass. For $m_{\scriptscriptstyle T}=600\, \text{GeV}$, it contributes with $28\%$ to the LO+PS cross section of the reaction $pp\rightarrow T\bar t+\bar T t\rightarrow t\bar t H$.

\acknowledgments

I am very thankful to P. Aurenche and J. Ph. Guillet for useful discussions and precious remarks. I acknowledge the hospitality of LAPTh where a portion of this work was done. 



\end{document}